\documentclass[twocolumn]{aastex63}

\usepackage[utf8]{inputenc}
\DeclareUnicodeCharacter{2212}{-}
\usepackage{graphicx}
\graphicspath{{./figures/}{./}}
\usepackage{amsmath}
\usepackage{afterpage}
\usepackage{enumitem}
\setenumerate{itemsep=0mm}
\hypersetup{    
  urlcolor        = {blue},
}
\usepackage{comment}
\usepackage{multirow}

\newcommand{\GK}[1]{#1}

\usepackage{rotating}

\usepackage{lineno}

\usepackage{soul} 
\usepackage{amsmath}
\usepackage{amssymb}
\usepackage{xspace}
\usepackage{xifthen}
\usepackage{eso-pic}

\definecolor{forestgreen}{HTML}{228B22}
\definecolor{urlblue}{HTML}{000000}

\newcommand{\response}[1]{{}}

\mathchardef\mhyphen="2D

\newlength{\dhatheight}

\newcommand{\code}[1]{\texttt{#1}\xspace}

\newcommand{\unit}[1]{\ensuremath{\mathrm{\,#1}}\xspace}

\newcommand{\Msolar}{\unit{M_\odot}}

\newcommand{\Mstar}{\unit{M_{*}}}

\newcommand{\e}{\unit{e^{-}}}

\newcommand{\tabref}[1]{Table~\ref{tab:#1}}

\newcommand{\bandvar}[2][]{%
  \ifthenelse{\isempty{#1}}{\var{#2}}{\var{#2\_#1}}%
}

\newcommand{\var}[1]{\ensuremath{\texttt{\MakeUppercase{#1}}}\xspace}

\providecommand\physrep{\ref@jnl{Phys.~Rep.}}%
\providecommand\apjs{\ref@jnl{ApJS}}%
\providecommand{\jcap}{\ref@jnl{JCAP}}%

\begin{document}

\title{Synthesizing Stellar Populations in South Pole Telescope Galaxy Clusters: I. Ages of Quiescent Member Galaxies at $0.3 < z < 1.4$}
\shorttitle{Ages of Quiescent Galaxies in SPT Galaxy Clusters}
\shortauthors{Khullar et al. 2021}


\author[0000-0002-3475-7648]{Gourav Khullar}
\affiliation{Department of Astronomy and Astrophysics, University of
Chicago, 5640 South Ellis Avenue, Chicago, IL 60637}
\affiliation{Kavli Institute for Cosmological Physics, University of
Chicago, 5640 South Ellis Avenue, Chicago, IL 60637}
\affiliation{Kavli Institute for Astrophysics \& Space Research, Massachusetts Institute of Technology, 77 Massachusetts Ave., Cambridge, MA 02139, USA}

\author[0000-0003-1074-4807]{Matthew B. Bayliss}
\affiliation{Department of Physics, University of Cincinnati, Cincinnati, OH 45221, USA}

\author[0000-0003-1370-5010]{Michael D. Gladders}
\affiliation{Department of Astronomy and Astrophysics, University of Chicago, 5640 South Ellis Avenue, Chicago, IL 60637}
\affiliation{Kavli Institute for Cosmological Physics, University of
Chicago, 5640 South Ellis Avenue, Chicago, IL 60637}

\author{Keunho J. Kim}
\affiliation{Department of Physics, University of Cincinnati, Cincinnati, OH 45221, USA}

\author[0000-0002-2238-2105]{Michael S Calzadilla}
\affiliation{Kavli Institute for Astrophysics \& Space Research, Massachusetts Institute of Technology, 77 Massachusetts Ave., Cambridge, MA 02139, USA}

\author{Veronica Strazzullo}
\affiliation{INAF – Osservatorio Astronomico di Trieste, Via Tiepolo 11, 34131 Trieste, Italy}
\affiliation{INAF – Osservatorio Astronomico di Brera, Via Brera 28, 20121 Milano, Via Bianchi 46, 23807 Merate, Italy}

\author{Lindsey E. Bleem}
\affiliation{Kavli Institute for Cosmological Physics, University of
Chicago, 5640 South Ellis Avenue, Chicago, IL 60637, USA}
\affiliation{Argonne National Laboratory, High-Energy Physics Division,9700 S. Cass Avenue, Argonne, IL 60439, USA}

\author[0000-0003-3266-2001]{Guillaume Mahler}
\affiliation{Institute for Computational Cosmology, Durham University, South Road, Durham DH1 3LE, UK}
\affiliation{Centre for Extragalactic Astronomy, Durham University, South Road, Durham DH1 3LE, UK}

\author{Michael McDonald}
\affiliation{Kavli Institute for Astrophysics \& Space Research, Massachusetts Institute of Technology, 77 Massachusetts Ave., Cambridge, MA 02139, USA}

\author[0000-0003-4175-571X]{Benjamin Floyd}
\affiliation{Department of Physics and Astronomy, University of Missouri—Kansas City, 5110 Rockhill Road, Kansas City, MO 64110, USA}

\author[0000-0003-2226-9169]{Christian L. Reichardt}
\affiliation{School of Physics, The University of Melbourne, Parkville, VIC 3010, Australia}

\author{Florian Ruppin}
\affiliation{Kavli Institute for Astrophysics \& Space Research, Massachusetts Institute of Technology, 77 Massachusetts Ave., Cambridge, MA 02139, USA}

\author{Alexandro Saro}
\affiliation{Astronomy Unit, Department of Physics, University of Trieste, via Tiepolo 11, I-34131 Trieste, Italy}
\affiliation{Institute for Fundamental Physics of the Universe, Via Beirut 2, 34014 Trieste, Italy}
\affiliation{Osservatorio Astronomico di Trieste, via G. B. Tiepolo 11, I-34143 Trieste, Italy}
\affiliation{National Institute for Nuclear Physics, Via Valerio 2, I-34127 Trieste, Italy}

\author[0000-0002-7559-0864]{Keren Sharon}
\affiliation{Department of Astronomy, University of Michigan, 1085 South University Drive, Ann Arbor, MI 48109, USA}

\author{Taweewat Somboonpanyakul}
\affiliation{Kavli Institute for Particle Astrophysics \& Cosmology (KIPAC), 452 Lomita Mall, Stanford, CA 94305}

\author[0000-0003-0973-4900]{Brian Stalder}
\affiliation{Vera C. Rubin Observatory Project Office, 950 N. Cherry Ave, Tucson, AZ 85719, USA}
\affiliation{Center for Astrophysics — Harvard \& Smithsonian, 60 Garden Street, Cambridge, MA 02138, USA
}

\author[0000-0002-2718-9996]{Antony A. Stark}
\affiliation{Center for Astrophysics | Harvard \& Smithsonian, 60 Garden St, Cambridge, MA 02138}

\email{gkhullar@uchicago.edu}

\submitjournal{ApJ}
\received{November 16th, 2021}

\begin{abstract}

Using stellar population synthesis models to infer star formation histories (SFHs), we analyse photometry and spectroscopy of a large sample of quiescent galaxies which are members of Sunyaev-Zel'dovich (SZ)-selected galaxy clusters across a wide range of redshifts. We calculate stellar masses and mass-weighted ages for 837 quiescent cluster members at $0.3 < z < 1.4$ using rest-frame optical spectra and the Python-based \texttt{Prospector} framework, from 61 clusters in the SPT-GMOS Spectroscopic Survey ($0.3 < z < 0.9$) and 3 clusters in the SPT Hi-z cluster sample ($1.25 < z < 1.4$). We analyse spectra of subpopulations divided into bins of redshift, stellar mass, cluster mass, and velocity-radius phase-space location, as well as by creating composite spectra of quiescent member galaxies. We find that quiescent galaxies in our dataset sample a diversity of SFHs, with a median formation redshift (corresponding to the lookback time from the redshift of observation to when a galaxy forms 50\% of its mass, t$_{50}$) of $z=2.8\pm0.5$, which is similar to or marginally higher than that of massive quiescent field and cluster galaxy studies. We also report median age-stellar mass relations for the full sample (age of the Universe at $t_{50}$ (Gyr) = $2.52 (\pm0.04) - 1.66 (\pm0.12)$ log$_{10}(M/10^{11}\Msolar)$) and recover downsizing trends across stellar mass; we find that massive galaxies in our cluster sample form on aggregate $\sim0.75$ Gyr earlier than lower mass galaxies. We also find marginally steeper age-mass relations at high redshifts, and report a bigger difference in formation redshifts across stellar mass for fixed environment, relative to formation redshifts across environment for fixed stellar mass.
\end{abstract}


\section{Introduction}
\label{sec:intro}

How and whether a given galaxy undertakes the path from initial star formation, to quenching, to passive evolution thereafter, is a fundamental question in the field of galaxy evolution. Studies that characterize galaxy mass assembly as a function of stellar content, halo mass, and environment are a path forward in both defining and solving the problem. Spectral energy distribution (SED) fitting and stellar population synthesis modeling originated as methods to study populations of elliptical galaxies with \cite{tinsley1976}. In the last several decades, with the extensive development of computational tools, photometry-based SED fitting has become a pivotal method to measure properties such as stellar masses, ages, and metallicities of a diverse population of galaxies, allowing us to study mass assembly in these systems.

This technique has been applied to a wide variety of spectroscopic, and in particular photometric, data across a range of galaxy populations that sample an abundance of intrinsic properties (e.g., star formation rate, stellar mass, metallicity, ages and environment). Recent multi-wavelength surveys have been successful in studying representative samples of quiescent galaxies in the field up to $z>3$ (e.g., \citealt{Heavens2000,Cimatti2004,Daddi2005,Gallazzi2005,gallazzi2014,Onodera2012,Onodera2015,Jorgensen2013,whitaker2013,fumagalli2016,pacifici2016}). These observations have confirmed that the number density of massive quiescent galaxies in the field has increased by an order of magnitude since z$\sim$2 \citep{Ilbert2013,muzzin2013,fumagalli2016}. Numerous studies also discuss both the timescales of cessation of star formation, and the likely processes responsible for quenching, noting that the efficacy of some of these processes is a strong function of environment (some recent works include \citealt{carnall2018,carnall2019,leja2019,tacchella2021}). Ram pressure stripping is thought to be effective in dense environments - i.e. the cores of galaxy clusters \citep{gunn1972,larson1980, balogh2000} — whereas strangulation of a galaxy's cold gas supply through a variety of possible mechanisms, resulting in a slow cessation of star-formation, is operative over a larger range of environmental densities \citep{peng2015}. Galaxy harassment — high speed dynamical encounters that are particularly common in the cluster environment — also likely plays a role, and may be particularly effective in driving the morphological transformation that accompanies the cessation of star formation in quenched systems (e.g., \citealt{moore1998}). Internal feedback processes, in particular active galactic nuclei (AGN)-feedback, are also thought to influence quenching \citep{dave2016,dave2019,nelson2019}.

As conducting observational longitudinal studies of galaxies — that track the evolution of galaxies through cosmic time — is impossible, studies resort to drawing conclusions from observations of different galaxies at different redshifts; this approach is challenging since galaxies sample a diverse set of star formation histories (SFHs). Moreover, recent work \citep{kelson2014,abramson2016} has shown that imprints of quenching are not necessarily distinguishable in the observations of quiescent galaxies. This makes an understanding of the evolutionary connection between galaxies across time difficult to elucidate in anything but the bulk statistical properties (e.g., luminosity or mass functions, color distributions, etc.). 

When investigating galaxy clusters, we have the opportunity to utilise the host cluster halo evolution - which is well described and understood from even dark-matter-only simulations (see \citealt{kravtsov2012} and references therein) - to connect the cluster galaxy populations in antecedent-descendant $clusters$ and hence construct a longitudinal sample of cluster galaxies. Galaxy clusters are unique environments with an abundance of observational constraints, 
and with a richness of passively evolving galaxies to study. In such analyses, one must carefully consider the effect of sample selection; for example, a fixed observational definition of quiescence applied at different redshifts results in some degree of progenitor bias \citep{vandokkum2001}.

Studies that analyze cluster galaxies (both as individual objects and in aggregate) at $z < 1$ suggest that massive galaxies in clusters form stars in an epoch of early and rapid star formation (at $z >\sim 3$), before quickly settling into a mode of quiescent evolution \citep{dressler1982,stanford1998,balogh1999,dressler2004,stanford2005,holden2005,mei2006}. Thus, observations of clusters at higher redshifts should sample an epoch where this star formation --- or at least its end stages --- is observed \textit{in situ}. Several studies of often small heterogeneous samples of galaxy clusters at $1 < z < 2$ have shown high star formation, AGN activity, and blue galaxy fraction compared with lower redshifts, as well as an evolving luminosity function \citep{hilton2009,tran2010,mancone2010,mancone2012,fassbender2011,snyder2012,brodwin2013,alberts2014a}. This is evidence that cluster galaxies are undergoing significant stellar mass assembly in this epoch, inviting further investigation into properties of member galaxies as well as the intra-cluster medium (ICM) at $z > 1$.

Studies have compared galaxy cluster environments with field galaxies to chart the role that these dense environments and deep gravitational potential wells play in the transition of galaxies from star-forming to quiescent \citep{balogh1999,ellingson2001,dressler2013,webb2020}. These studies characterize ages and SFHs of massive galaxies -- both quiescent and star forming -- and infer quenched fractions of galaxies in these environments. 

Despite these successes, some challenges remain, especially constructing cluster samples across a wide range of redshifts. This is due to the following reasons. First, optical, IR and X-ray fluxes -- which are observational tracers of galaxy clusters -- become progressively more difficult to measure at high redshift due to cosmological dimming \citep{bohringer2013,bartalucci2018}. Second, to conduct evolutionary studies and characterize the precursors of lower-redshift clusters, we need to study the appropriate antecedents of lower-redshift massive clusters — which are high-redshift lower-mass systems. This is a non-trivial sample to build; $z>1$ systems measured with these observations are few in number \citep{2005ApJ...634L.129S,2012ApJ...753..164S,2014ApJS..213...25S,2006ApJ...651..791B,2011ApJ...732...33B,2006ApJ...639..816E,2006astro.ph..4289W,2008ApJ...684..905E,2009ApJ...698.1934M,2010ApJ...716.1503P,2010ApJ...711.1185D,2011MSAIS..17...66S,2012ApJ...759L..23G,2012ApJ...756..115Z,2015ApJ...812L..40G,2017MNRAS.470.4168B,2017ApJ...844...78P}. Third, a challenge with optical and IR cluster surveys is whether the selection of galaxy clusters based on member galaxy properties systematically affects the studies of the said galaxies, e.g., while red-sequence selection of clusters has proven extremely fruitful for finding clusters and groups across a broad range of mass and redshift, it remains a concern whether this selection biases our understanding of quiescent (i.e., red-sequence) cluster galaxies, particularly at higher redshifts. \GK{By comparison, an ICM-selected cluster sample — a mass-limited sample where the limit doesn't evolve significantly with redshift — is likely to be less biased for galaxy evolution studies in clusters.} Clusters discovered via the Sunyaev-Zel'dovich (SZ) effect with the South Pole Telescope (SPT, \citealt{carlstrom2002}) and the Atacama Cosmology Telescope (ACT, \citealt{sifon2016}) provide a nearly mass-limited sample of clusters with a redshift-independent mass threshold set by instrument sensitivity. Recent SZ-based galaxy cluster searches from SPT have revealed new samples of galaxy clusters at z$>\sim$1 \citep{bleem2015,bleem2020,huang2020}, extending the viability of SZ-cluster studies to redshift as distant as any other sample. These samples are now large enough to be a compelling resource for cluster galaxy evolution studies \citep{brodwin2010,stalder2013,mcdonald2013,ruel2014,bayliss2016,mcdonald2017,khullar2019}. 

Another challenge in conducting SED-based studies is tied to how reliably we can interpret physical properties inferred from photometric observations compared with spectroscopic data. Studies relying on photometry alone are subject to many challenges, such as the age-metallicity-dust degeneracy \citep{worthey1994,charlot1999}. 

\GK{While large photometric samples of galaxy populations exist ranging from the present epoch to z $\sim$ 2-3 for L*-type (and fainter) galaxies, only in the last two decades have statistical studies of galaxies with spectra and SED fitting been conducted, particularly on quiescent galaxies at intermediate and high redshifts in the field (e.g., \citealt{juneau2005,gobat2008,2010ApJ...711.1185D,moresco2010,moresco2013,choi2014,dressler2016,belli2019,carnall2019c,estradacarpenter2020,tacchella2021}) and clusters (\citealt{sanchezblazquez2009,muzzin2012,jorgensen2017,webb2020})}. Moreover, recent complex numerical simulations have been able to reproduce many physical conditions of galaxies and demonstrate hierarchical structure formation (e.g., \citealt{springel2005,dave2019,nelson2019}), as well as approximately infer the parameters of evolution required to connect high-redshift galaxies to low-redshift descendants. Robust analyses of spectroscopic data can aid in comparison with these simulations, as well as inform simulations of galaxies at high-redshifts ($z>1$). 

We conduct here a study of stellar populations in quiescent cluster galaxies, and the influence of a systematically-selected cluster environment on the evolution of these member galaxies. We aim to answer the following questions:

\begin{enumerate}
    \item On what timescales did galaxies that end up in galaxy clusters form their stars?
    \item How does the cluster environment and location of a given galaxy within the cluster affect this timescale?
    \item  While studying these properties, does the galaxy cluster selection method matter?
\end{enumerate}

We use 63 SZ-selected clusters from the SPT-SZ Survey (\citealt{bleem2015}, hereafter LB15) across 0.3 $< z <$ 1.4 with extensive spectroscopy \citep{bayliss2016, khullar2019}, and characterize 837 quiescent galaxies spectrophotometrically to address the above questions. Because SZ cluster samples can reach lower mass thresholds at high redshifts, this SZ cluster sample connects high-redshift lower-mass antecedent clusters to low-redshift higher-mass descendents. Here, we study the evolution of quiescent galaxy population across redshift in this antecedent-descendent cluster sample.

This paper is organized as follows. Section \ref{sec:data} lays out the photometric and spectroscopic data used in this work, Section \ref{sec:construct_sample} describes the quiescent galaxy sample construction, and Section \ref{sec:analysis} describes the methods used in our analysis. Section \ref{sec:results} and Section \ref{sec:ages} describe mass-weighted ages and formation redshifts for individual galaxies and subpopulations binned by various properties. We discuss some challenges in this work and future directions in Section \ref{sec:challenges}. Finally, we summarise our work in Section \ref{sec:conc}.

Magnitudes have been calibrated with respect to the AB photometric system. The fiducial cosmology model used for all distance measurements as well as other cosmological values assumes a standard flat cold dark matter universe with a cosmological constant $(\Lambda$CDM), corresponding to WMAP9 observations \citep{Hinshaw_2013}. All Sunyaev-Zel'dovich (SZ) significance-based masses from LB15 are reported in terms of  M$_{500c,SZ}$ i.e. the SZ mass within R$_{500c}$, defined as the radius within which the mean density $\rho$ is 500 times the critical density $\rho_{c}$ of the universe.

\section{Observational Data}

\label{sec:data}

In order to perform a comparative analysis on individual and aggregate stellar populations of quiescent member galaxies in our massive galaxy cluster sample, we combine the low-redshift SPT-GMOS cluster spectroscopic sample (from \citealt{bayliss2016,bayliss2017}, hereafter B16 and B17) with spectra from the SPT Hi-z galaxy cluster sample (\citealt{khullar2019}, hereafter K19), to give us a sample of 63 galaxy clusters from $0.3 < z < 1.4$. 

\begin{figure*}[htb!]
\centering
\includegraphics[width=1.0\textwidth]{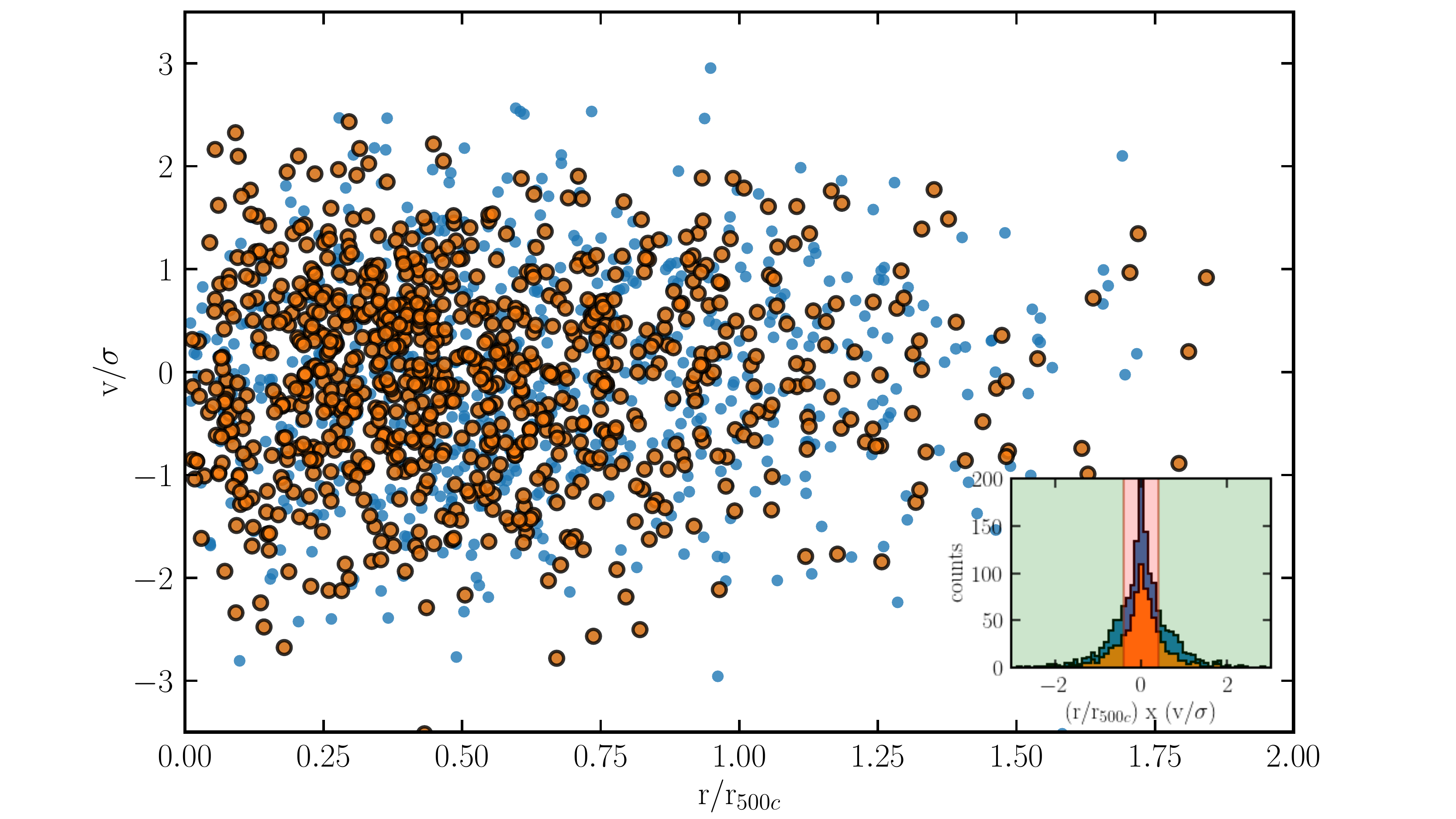}
\caption{Normalized proper velocity vs. normalized distance of member galaxies in the sample from the nominal cluster SZ center. Velocities are normalized to the velocity dispersion of the galaxy cluster. Orange points correspond to quiescent galaxies in the sample, while blue points represent non-quiescent galaxies. (Inset) A histogram of the distribution of phase-space location (or proxy for `infall time') for all galaxies (blue) and quiescent galaxies (orange), defined as p = r$\rm_{projected}$/$r\rm_{500c} \times$ $v\rm_{peculiar}/\sigma_{v}$. The shaded red region corresponds to galaxies in the `early+mixed infall' subpopulation i.e. p$<0.4$.}
\label{fig:phasespace_sgal}
\end{figure*}

\begin{figure}[htb!]
\centering
\includegraphics[width=0.47\textwidth]{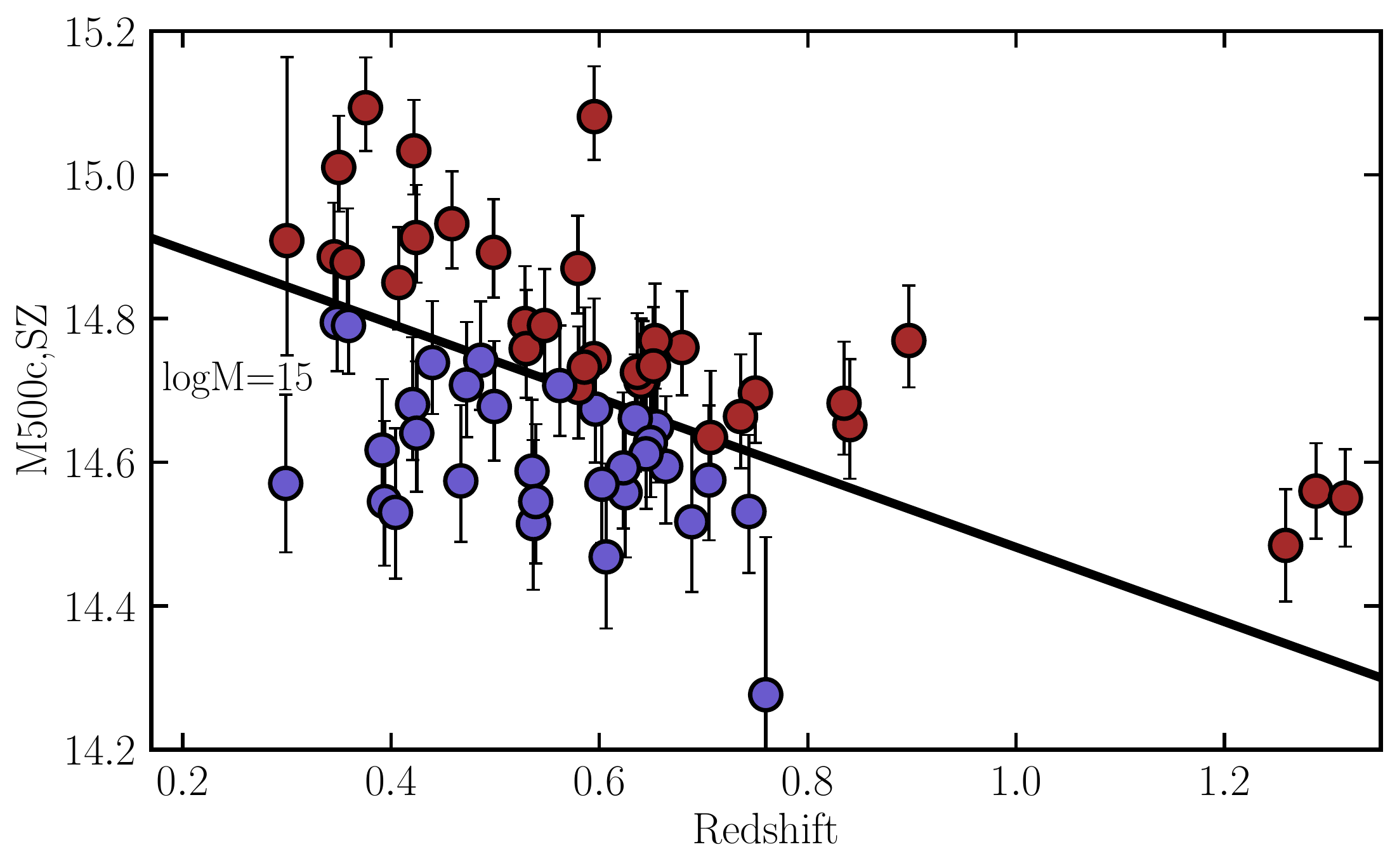}
\caption{log(M$_{500c,SZ}$/\Msolar) vs redshift for 64 clusters in the High-z and Low-z samples. The black line demarcates the evolutionary path of galaxy clusters to final cluster mass at redshift 0 of log(M$_{500c,SZ, z=0}/\Msolar$) = 15. The cluster sample is divided into two groups based on this demarcation, to facilitate a descendant-antecedent analysis of member galaxies \citep{Fakhouri2010} — blue (red) points are clusters with final descendant cluster mass of log(M$_{500c,SZ, z=0}/\Msolar$) $<15$ ( $>15$).}
\label{fig:m_finaldesc_z}
\end{figure}

For all spectra considered in this work, we ensure that spectroscopic features being used to characterize SFHs are consistent across redshift and surveys in the rest frame. In this study, we use all spectra across galaxies in the rest frame wavelength range 3710-4120$\AA$. To classify galaxies at the catalog level, and isolate the passively evolving subset, we use rest-frame [OII] $\lambda$ 3727,3729\AA\ doublet emission lines (blended here) and the D4000 spectral index (ratio of the spectral flux blueward and redward of the 4000\AA\ break); these rest-frame optical signatures in spectral data are age indicators of stellar populations and are used for making quiescent galaxy cuts in our data. Numerous spectral features such as the CN molecular band, Ca II H$\&$K $\lambda$ 3968,3934\AA\, H$\delta$, H9, H10 and H11 absorption features are also present in this wavelength range, and contribute to the full spectrophotometric SED fitting. 

For all galaxies in our sample, we note environment- and cluster-specific properties, namely their velocity-radius phase space locations and the final descendant mass of the host galaxy cluster. We label each galaxy with its location in their proper velocities vs. normalized distance from cluster center space (see \citealt{noble13}, and Figure \ref{fig:phasespace_sgal}) to assign a proxy for galaxy `infall time', in order to compare galaxies that are at different stages in their trajectory after infalling into their corresponding galaxy cluster. We also assign M$_{500c,SZ, z=0}$, which is the inferred final descendant cluster mass M$_{500c,SZ}$ at redshift $z=0$ (assuming a halo mass growth history, \citealt{Fakhouri2010}), and label our galaxy sample with their membership in clusters with log(M$_{500c,SZ, z=0}/\Msolar$) greater or lesser than 15 (see Figure \ref{fig:m_finaldesc_z}). For further details, we direct the reader to Section \ref{sec:stack_description}.

\subsection{High-z Cluster Spectroscopy: $1.2 < z < 1.4$}

The high redshift cluster sample in this work is from K19, which spectroscopically confirmed five galaxy clusters at $1.25 <$ \textit{z} $< 1.5$. This sample, comprising 5 of the 8 most massive SPT-SZ clusters at $z>1.2$, was assembled for a deep \textit{Chandra} X-ray Observatory X-ray imaging program \citep{mcdonald2017}. We identify 10 of the 44 member galaxies characterized in K19 as passive (see Section \ref{sec:construct_sample}), and include them in this work for analysis on individual spectra, as well as to construct a higher signal-to-noise ratio (S/N) composite spectrum for the redshift bin $1.2 < z < 1.4$ (see observation details in Section 2 of K19). Note that only 3 $z>1.2$ clusters have been shown in Figure \ref{fig:m_finaldesc_z}) - the 10 quiescent galaxies are member galaxies of these clusters.

The spectra in this sample typically cover the wavelength range 7500-10000 \AA\ in the observed frame, and the rest-frame range 3710-4120 \AA\ is common to all spectra across $1.2 < z < 1.4$; the low-redshift spectra sample's rest-frame wavelength range is matched appropriately. 
This dataset has low S/N observations, and these spectra are dominated in some spectral ranges by sky background noise associated with sky-subtraction residuals, an artifact of both the data quality and limitations of the reduction process. Due to the low S/N of the dataset, getting robust constraints on stellar population properties is difficult (see Section 4.2 in K19 for a discussion of constraining redshift uncertainties for these spectra). We present results from SED fitting of individual galaxies with this caveat in mind, but also lean on results from a stacked quiescent galaxy spectrum comprising 10 quiescent galaxies in the sample  cut on [OII] and D4000 identically to the lower-z sample (see Section \ref{sec:construct_sample}). This cut is more restrictive than that used in K19, in which an initial stacked spectral analysis was presented.

\subsection{Low-z Cluster Spectroscopy: $0.3 < z < 0.9$}

The South Pole Telescope - GMOS Spectroscopic Survey cluster sample (SPT-GMOS; from B16 and B17) is a spectroscopic study of 62 galaxy clusters ($0.3 < z < 1.1$) from the SPT-SZ Survey cluster sample. The full sample of spectra contains 2243 galaxies including 1579 galaxy cluster members, confirmed in B16 and B17 via interloper exclusion and a velocity-radius phase space analysis (e.g., see \citealt{rhee2017,pasquali2019}and references therein). The data set used here consists of 1D flux calibrated spectra, redshifts, positions, velocity dispersions, equivalent widths of spectral features([OII],H$\delta$) and the spectral index D4000. This sample contains one cluster between $0.9<z<1.1$ — SPT-CL J0356-5337 at $z=1.03$ — with only 8 spectra of interest. We remove this cluster from consideration in this study; our analysis would require $0.9<z<1.1$ to be a single cluster redshift bin with 8 quiescent galaxies, which can significantly bias the results inferred from this bin.

\begin{figure}[htb!]
\centering
\includegraphics[width=0.5\textwidth]{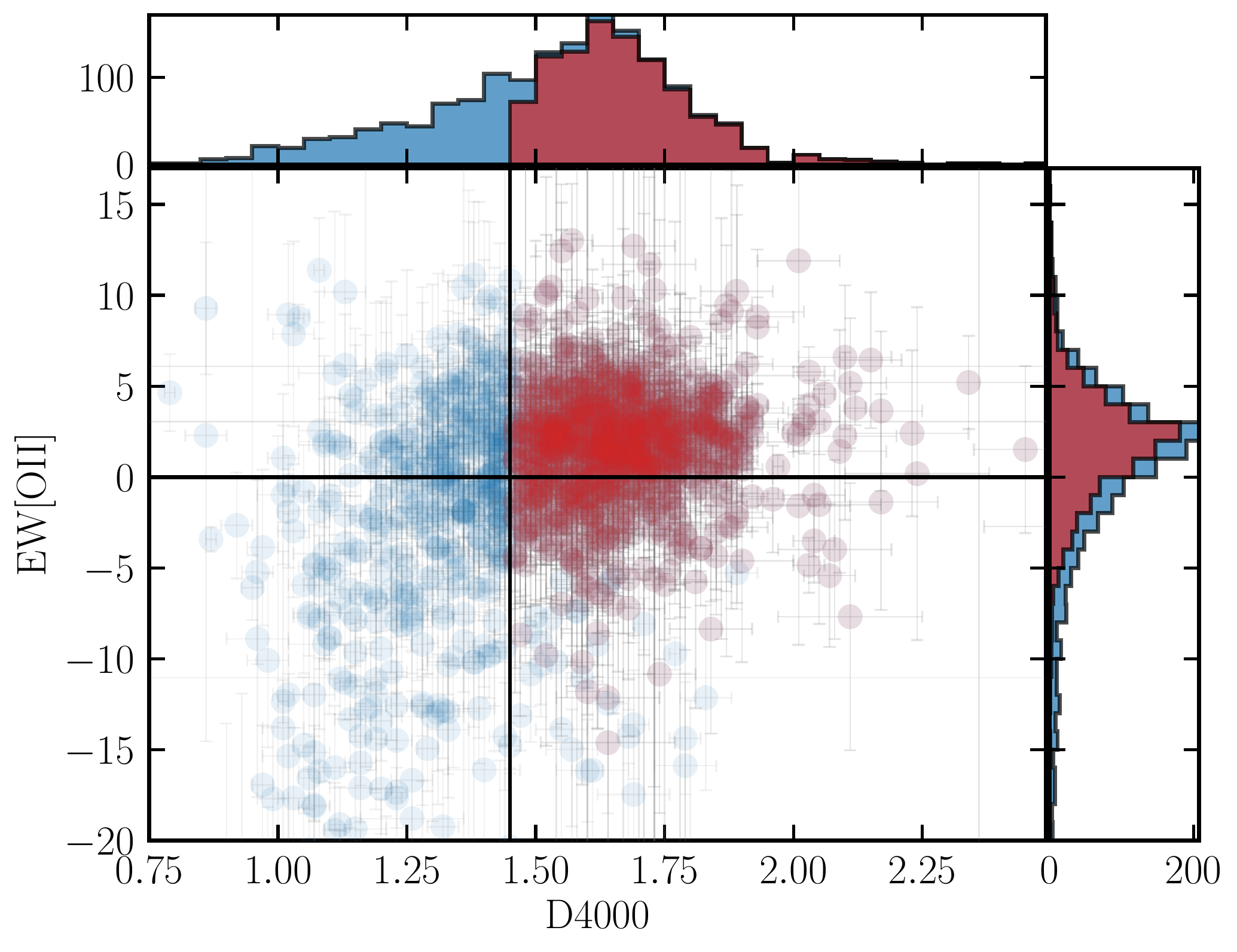}  
\caption{Distribution of equivalent width EW[OII] vs spectral index D4000 for sample galaxies (blue) and galaxies classified as quiescent in this work (red). EW[OII] vs D4000 is used here as an indicator for quiescent vs non-quiescent (actively star-forming, starburst or post-starburst) galaxies. Horizontal and vertical lines demarcating regions in the equivalent-width phase spaces are taken from \cite{1983ApJ...273..105B} and \cite{balogh1999} — quiescent galaxies have D4000 $>1.45$ and no detection of an [OII] emission feature at $>2\sigma$. We also test a more probabilistic cut for D4000 (i.e. D4000$>1.45$ at $>2\sigma$), which does not significantly impact our results.}
\label{fig:d4000_o2}
\end{figure}

\subsection{Photometry}

The flux calibration of the spectra used here suffers from the usual limits of multi-object spectroscopy of extended sources: aperture losses that are a complex function of observing conditions, source morphology and slit-mask details. Neither of the spectrographs — Gemini/GMOS and Magellan/LDSS3 — that contribute to these data have atmospheric dispersion correctors, and hence the flux calibration has potential wavelength dependencies that result from observing multi-object slit-masks at generally non-parallactic angles and a range of airmasses.

We use $griz$ photometry for the purpose of doing joint spectrophotometric SED fitting such that flux calibration is a fitted nuisance parameter in our analysis. This correction allows us to calculate robust stellar masses for member galaxies in our cluster sample, which is a key property to characterize stellar populations. These stellar masses aid in making completeness cuts, as well as in assigning bins for stacking. The methodology, SED model definitions, and the analysis are described in Section 3.

We use available photometry for SPT Cluster galaxies used in B16 and B17, taken from a pool of optical imaging data used for SPT-SZ cluster confirmation and follow-up (\citealt{high2010,song2012}, LB15). This contains 1-4 band photometry ($griz$) for 60\% of member galaxies in our sample. To increase the number of galaxies for which at least one photometric data point is available (and hence allowing us to flux calibrate the spectra to the photometry and calculate robust stellar masses), we use additional photometry from the Parallel Imager for Southern Cosmology Observations (PISCO; \citealt{stalder2014}) catalog described in \citealt{bleem2020} (uniform depth $griz$ imaging data for over 500 SPT-selected clusters and cluster candidates). In summation, of a total of 1251 member galaxies, 978 quiescent galaxies in our sample have photometric data to supplement spectroscopic analysis.


\begin{figure*}[htb!]
\centering
\includegraphics[width=1.0\textwidth]{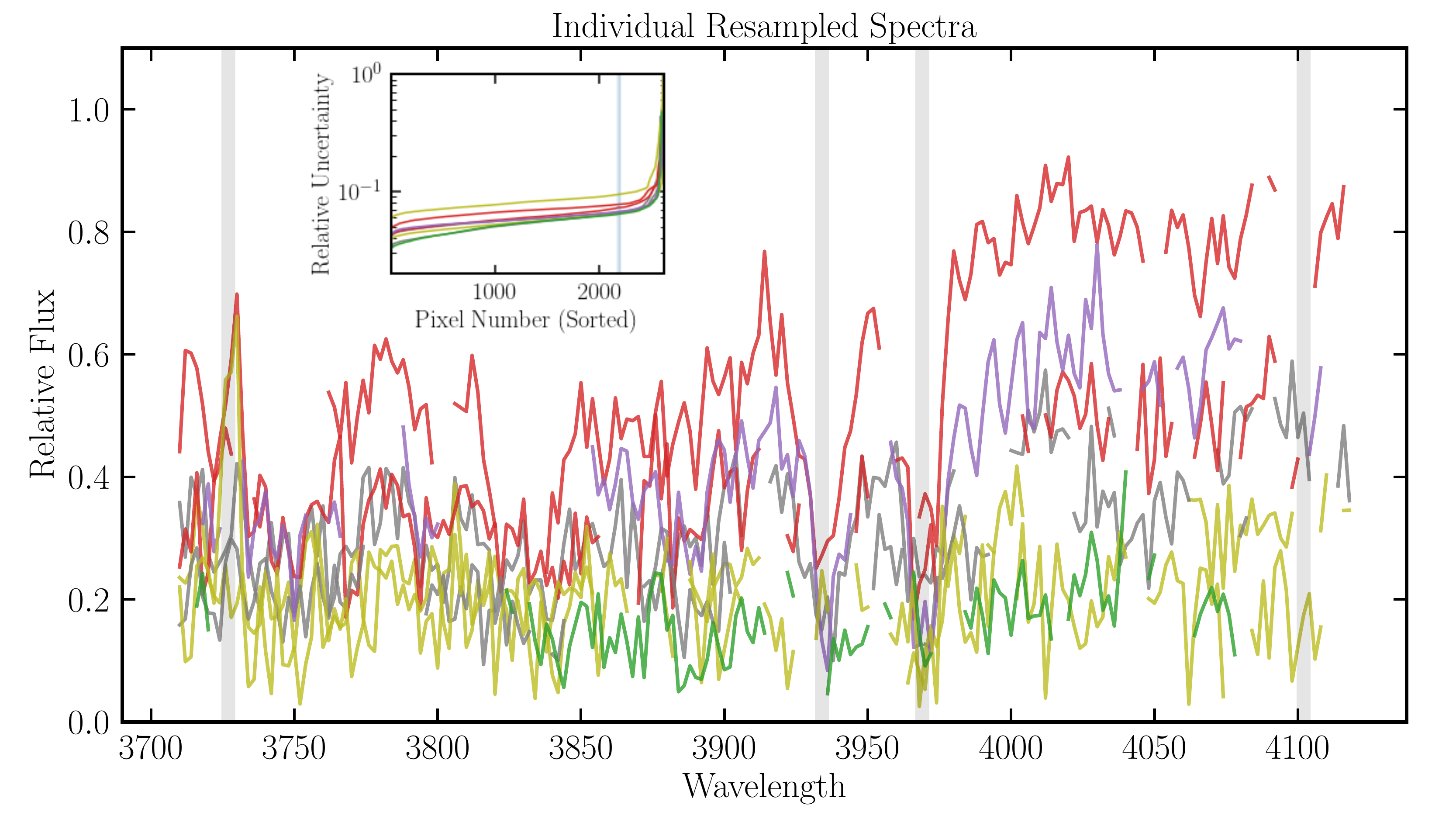}
\caption{Noise-masked and resampled rest-frame spectra for all observed galaxies, prior to signal-to-noise, quiescent galaxy and stellar mass selection, in the cluster SPT-CL J0013-4906 at z=0.41 (colors represent spectra from different member galaxies). Shaded gray regions represent spectral features of interest in SED fitting (from left to right) — [OII] $\lambda$3727, 3729\AA\ doublet, Ca II K$\&$H $\lambda$ 3934,3968\AA\, and H$\delta$ at 4102$\AA$. [OII] and 4000\AA\ break are used here to make quiescent galaxy cuts. (Inset) Relative sorted flux uncertainties per pixel for the same spectra in SPT-CL J0013-4906. The blue vertical lines denote the error threshold (84th percentile) above which spectral pixels are masked, followed by resampling.}
\label{fig:stack_example}
\end{figure*}

\section{Constructing a sample of quiescent galaxies}
\label{sec:construct_sample}

\subsection{Spectroscopic Target Selection}

B16 and B17 placed slits on targets in the SPT-GMOS survey as follows: the highest priority was assigned to candidate brightest cluster galaxies (BCGs), followed by likely cluster member galaxies that were selected from the red sequence (identified as an overdensity in color-magnitude and color-color space) down to an absolute magnitude limit of $M^{*}+1$ (see Page 14 in B17 for a detailed description). Within this red-sequence selected galaxy sample, no magnitude prioritization was used, so the slits should randomly sample the red-sequence galaxy population down to the chosen limit. A similar procedure was followed for the high-z cluster galaxy sample in K19, though we note that the effective limiting absolute magnitude is brighter in these distant clusters. 

\GK{Based on these selection criteria for multi-object slits and completeness of observations in both the SPT-GMOS survey (B16, B17) and the SPT Hi-z survey (K19), we expect the member galaxy spectra to be least biased and most representative for red or quiescent galaxies, with brighter galaxies in a given cluster being observed with a high signal-to-noise ratio (SNR). We emphasize here that the B16 cluster galaxy sample is not mass complete sample, but a representative sample of quiescent member galaxies in SPT galaxy clusters. For more details, see Section \ref{sec:incomplete}.} 

For our SED analysis, we isolate this representative sample of galaxies by the following cuts in catalog space.
\subsubsection{Equivalent Width and Signal-to-Noise ratio}

B16 and B17 make informative cuts on quiescent,actively star-forming and post-starburst galaxies using physically motivated spectral indices (we invite the reader to view Table 3 from B17, and \citealt{balogh1999} for more details). For our main analysis, we use the B16 and \cite{balogh1999} data cuts to identify quiescent galaxies, as galaxies with no [OII] emission feature at $>2\sigma$, and D4000 $>$ 1.45.

We test the distribution of galaxies categorized as quiescent with stricter cuts — assuming Gaussian uncertainties on each EW, if a galaxy's EW (e.g., EW([OII]) is above or within 1$\sigma$ of the ``passive" threshold, we label the galaxy passive; the resulting sample is not significantly different from galaxies selected via the B16 quiescent galaxy cuts on the equivalent widths. See Figure \ref{fig:d4000_o2} for equivalent width and spectral index cuts implemented in this work.

We also calculate the mean SNR per pixel across each galaxy spectrum in the wavelength range 3710-4120, and remove galaxies with SNR$<$5 from our sample, since these are mostly galaxies without any robustly detected spectral features, and/or uncertainties that are non-Gaussian and dominated by systematic uncertainties due to sky subtraction. 

\subsection{Excluding Brightest Cluster Galaxies}

Our analysis focuses on the evolution and build-up of stars in the quiescent galaxy population in massive galaxy clusters, and it is important to note the role BCGs may play in biasing this analysis. BCGs are objects evolving through complex pathways near/at the center of the gravitational potential in clusters, and at the hub of merging and feedback activity in the cluster \citep{rawle2012,webb2015,mcdonald2016,pintoscastro2019}. We treat this population of galaxies as unique, dissociated from the quiescent galaxy analysis that is central to this paper.  

Extensive follow-up optical/IR photometry and X-ray observations for the clusters in this work was undertaken since the SPT-GMOS Spectroscopic Survey was published. These allow robust identifications of BCGs, via X-ray and IR peak/centroid characterization, through Chandra and Spitzer data respectively (Calzadilla et al. in prep). Table 3 in B16 provides a list of candidate BCGs for the SPT-GMOS sample. These are galaxies selected on the basis of their optical/IR flux and the projected spatial location in the cluster. We find that only 36 of these galaxies correspond to BCGs identified via X-ray and IR peak/centroid characterization through \textit{Chandra} and \textit{Spitzer} data respectively. 3 BCGs in the sample of high-z clusters ($z>1.25$) are reported in K19. These 39 galaxies are removed from the galaxy sample characterized here. Note that the number of BCGs excluded is less than the number of clusters, with the understanding that not all BCGs were spectroscopically observed/confirmed in the surveys considered here, or passed our data cuts. 

\subsubsection{Stellar Mass}

We calculate SED fitting-based stellar masses for all galaxies with available spectrophotometry (see Section 4 for model and analysis details). For galaxies passing a nominal ``quiescent galaxy" threshold (lack of [OII] emission, D4000 $> 1.45$), we adopt a homogenous cut (across redshift) at M $> 2\times$10$^{10}$ \Msolar, that ensures a uniform distribution of stellar masses across redshift bins; this removes a further 102 low-mass galaxies from the sample. See Section \ref{sec:incomplete} for more details.


\newcommand{\arccaption}{\code{Prospector} Analysis: Free Parameters in SED Model A}
\begin{deluxetable*}{l cccc}
\tablecolumns{4}
\tablewidth{0pt}
\tablecaption{\arccaption}
\tablehead{Parameter & Description & Priors}
\startdata
$M_{\rm total} (M_{\odot})$ & Total stellar mass formed & Log$_{10}$ Uniform: [10$^{9}$, 10$^{13}$] \\
$z$ & Observed Redshift (Mean Redshift from B16 and K19) & TopHat: [$z-0.002$, $z+0.002$]\\
$\log(Z/Z_{\odot})$ & Stellar metallicity in units of $\log(Z/Z_{\odot})$ & Clipped Normal: mean=0.0, $\sigma$=0.3, range=[-2, 0.5]$^{*}$\\
$t_{age}$ & Age of Galaxy & TopHat: [0, Age(Universe) at z$_{obs}$]\\
$\tau$ & e-folding time of SFH (Gyr) & Log$_{10}$ Uniform: [0.01, 3.0] \\
$\mathrm{spec}_{\rm norm}^{**}$ & Factor by which to scale the spectrum to match photometry & TopHat: [0.1, 3.0] \\
$\sigma_{v}$ & Velocity smoothing (km s$^{-1}$) & TopHat: [150.0, 500.0] \\
$(p1,p2,p3)^{**}$ & Continuum Calibration Polynomial (Chebyshev) & TopHat: n=3: [-0.2/(n + 1), 0.2/(n + 1)] \\
\enddata
{\footnotesize \tablecomments{$*$ Mean, $\sigma$ and range of the clipped normal priors based on the Mass-Metallicity relation (MZR) from Gallazzi et al. 2005.\\
$**$ Considered as nuisance parameters.}}
\label{table:sed_arc}
\end{deluxetable*}


\section{Methods and Analysis}\label{sec:analysis}

\subsection{Data Preparation}

\subsubsection{Masking 1D spectral pixels with unreliable noise properties}

In this work, we incorporate data from multiple instruments with a wide variety of operational parameters e.g., type of grism, observational seeing, etc. Moreover, the 1D spectra in our dataset contain pixels with significant sky subtraction residuals. In almost all cases, these pixels are represented by high noise/uncertainty which is ascribed as Gaussian, which may not be a robust assumption, especially for low SNR spectra observed from high-redshift galaxies, in multi-object slit observations where either the slit roughness or saturated sky contributes to poor sky subtraction. See work such as K19 and the Gemini Deep-Deep Survey \citep{abraham2004} for details on artefacts and mitigation strategies.

We mask these pixels in 1D spectra to prevent them from being taken into account in our SED analysis with the following framework. For each galaxy, we sort all uncertainties in increasing order, and attempt to characterise the knee of the uncertainty array, i.e. the value at which the uncertainty increases rapidly. For the majority of 1D galaxy spectra, this transition is captured by $\sim$ 84th percentile pixel in the uncertainty array (see inset of Figure \ref{fig:stack_example}). We mask all pixels between 84th-100th percentile of the sorted uncertainty array, as well as adjacent pixels, to eliminate pixels for which the uncertainty is large and poorly characterized.

\subsubsection{Resampling}

We resample 1D spectra and corresponding uncertainties to a common rest-frame wavelength grid, which facilitates both individual and stacked analysis. This is especially important since our spectra sample a wide redshift range, and as in any stellar population synthesis analysis it is crucial to avoid biases associated with non-uniform sampling of absorption line features (e.g., \citealt{leja2019b}). As mentioned in Section 2, the wavelength range common to all spectra in our sample is approximately 3710-4120\AA\ (rest frame). The resampling is performed on each spectrum using SPECTRES \citep{carnall2017b}, to a common wavelength range (3710-4120\AA\ rest frame) at 2\AA/pixel, coarser than both GMOS and LDSS3 1D spectral sampling. SPECTRES preserves integrated flux, and propagates uncertainties by calculating the covariance matrix for the newly binned/sampled spectra.

See Figure \ref{fig:stack_example} for examples of masked and resampled spectra from a representative cluster in our sample, SPT-CL J0013-4906 at z=0.41.


\subsection{SED fitting - Spectrophotometry of individual galaxies}\label{sec:sedfit_desc}

To characterise physical properties of member galaxies in our sample, we perform SED fitting to our spectrophotometry using the Markov Chain Monte Carlo (MCMC)-based stellar population synthesis (SPS) and parameter inference code, \code{Prospector} \citep{johnson21}. \code{Prospector} is based on the \code{Python-FSPS} framework, with the MILES stellar spectral library and the MIST set of isochrones
\citep{2010ApJ...712..833C,2017ApJ...837..170L,2013PASP..125..306F,2011A&A...532A..95F,Choi_2016}.

To test the robustness of our parameter inference and the model-dependence of physical properties, we fit our data with a fiducial model (Model A), and a minimalist model (Model B, see Appendix C). In this work, we specifically use two parametric SFHs — \GK{a delayed exponentially declining SFH (delayed-tau), and a single burst.} These models incorporate physical priors seen in studies of quiescent galaxies — low specific star formation rates (sSFRs), and a lack of rising star formation in galaxies (e.g., see \citealt{belli2019}).

\begin{itemize}
    \item{\textbf{Model A}: In this (fiducial) model, we fit as free parameters the total stellar mass formed (\Mstar), the stellar metallicity (log$(Z/Z_{sol})$), a delayed exponentially declining SFH, with age (t$_{age}$) and e-folding time ($\tau)$, and an internal smoothing parameter ($\sigma_{v}$ (km s$^{-1}$), to account for the contribution of Doppler broadening by stellar velocities, and resolution of the model libraries). The SFH, defined as the star formation rate as a function of time, is given by:
\begin{gather}
\operatorname{SFR}(t, \tau) \propto t / \tau * e^{-t / \tau}
\end{gather}

    To remove continuum calibration residuals (e.g., related to spectral response, flat fielding) from the spectra, we fit for a spectrophotometric calibration polynomial (a third-order Chebyshev polynomial; see \citealt{leja2019b,webb2020}).}

    \item{\textbf{\textit{Model B}}: Historically, simple stellar population (SSP) models with instantaneous episodes of star formation have been employed to characterise star formation in early-type galaxies. While not physical, such simple burst models are often assumed to be a proxy for a passively evolving stellar population when observed sufficiently long after a star forming episode. In order to compare our work with prior studies, we also fit a minimally complex model, comprising as free parameters the total stellar mass formed (\Mstar) and a single burst age t$_{age}$ that accounts for all the mass formed. We fix the metallicity to log$(Z/Z_{\odot})$ = 0.0 (solar metallicity), and treat the internal velocity smoothing as a fixed parameter at $\sigma_{v}$=280 km s$^{-1}$.} See analysis and results for Model B in Appendix A.
\end{itemize}

Both models assume a Kroupa IMF \citep{kroupa2001}, and no dust attenuation. Nebular continuum and line emission are turned off, as these are not expected to have significant contributions in fluxes of quiescent galaxies. Moreover, the flux normalization in our spectra is uncertain in practice and suffers from aperture losses; we account for this by including a nuisance parameter $spec_{norm}$ (spectrum normalization factor), to capture this. We do the fitting in observed wavelength space, and fit for a redshift parameter with narrow priors to capture uncertainties in the measured redshift.

\newcommand{\bincaption}{Binning Criteria and Description}
\begin{deluxetable*}{|l |c|c|c}
\tablecolumns{3}
\tablewidth{0pt}
\tablecaption{\bincaption}
\tablehead{Bin Description & No. of Bins & Criterion}
\startdata
Observed Redshift, z & 4 & $0.29<z<0.45 | 0.45<z<0.61 | 0.61<z<0.91 | 1.2<z<1.5$ \\
Stellar Mass, M$_{*}$ & 2 & $10.3<$logM$<10.9 | 10.9<$ logM $<12.1$\\
Final descendant cluster mass, log(M$_{500c,SZ, z=0}/\Msolar$) & 2 & logM$_{500c,finaldesc}$  $>15$ $|$ logM$_{500c,finaldesc}$ $<15$\\
Phase-space location, p = r$_{projected}/r_{500c}$ x $v_{peculiar}/\sigma_{v}$ & 2 & Early+Mixed infall:$p<0.4 |$ Late infall:$p>0.4$ \\
\enddata
\label{table:bins}
\end{deluxetable*}

There is considerable evidence shown in the literature that the prior probability densities assumed for the parameters related to the SFHs significantly impact the inferred parameter values; a linearly uniform prior in $\tau$ imposes a peaked and more informed prior probability density on the specific star formation rate (sSFR, the parameter of interest when fitting an SFH; see Figure 2 in \citealt{carnall2019}). Thus, the e-folding time parameter $\tau$ is constrained by fitting with a uniform prior in log-space, which is seen to be a less informative prior in sSFR. We implement uniform priors for the age parameter from 0 Gyr to the age of the Universe at the epoch of observation. 
The log$(Z/Z_{\odot})$ parameter is sampled with a Gaussian prior, clipped at -2.0 and 0.2. These bounds are limited by the extent of metallicity sampling in the MILES and MIST model libraries. The mean and $\sigma$ of this Gaussian are based on the Mass-Metallicity relation (MZR) from \cite{Gallazzi2005}; please see Appendix A for more discussion.

Within \code{Prospector}, we use \code{emcee} \citep{2013PASP..125..306F} to sample the posterior distribution of free parameters in each model, where burn-in, number of walkers, and number of iterations are selected iteratively, until convergence is seen to be reached (via visual confirmation) in the traces/steps of 32 randomly sampled walkers.

The details of these models, the model parameter definitions, and priors used here are laid out in Table \ref{table:sed_arc}. See Section \ref{sec:sfh_gal} for a discussion of analyses using the inferred SED parameters.

\subsection{Binning and stacking quiescent galaxy spectra} \label{sec:stack_description} 

To demonstrate aggregate properties of galaxies in our sample, we calculate median properties for different sub-populations of galaxies, and we perform stacking analyses on these same sub-populations. 

Before considering which galaxy spectra to stack and the implementation of a robust algorithm to do so, it is worth noting that stacking can result in biased inferences of galaxy properties, especially in scenarios where there is a highly non-linear correlation between spectral flux and the said property (e.g., metallicity evolution does not scale linearly with flux in any part of a typical galaxy spectrum). Thus, stacking should be considered with appropriate caution. That being said, the highest-redshift galaxies (at $z>1.2$) have severe sky subtraction residuals with non-Gaussian and ill-measured uncertainties, which do not allow us to reliably interpret these spectra via individual galaxy SED fitting only. Moreover, the galaxies in this subpopulation are at different redshifts between $1.2<z<1.4$; each individual spectrum is impacted at different rest wavelengths by skylines, allowing the stacking to ``fill in" much of the gaps. Therefore, to boost SNR as well as wavelength coverage, stacking provides us with an aggregate indication of galaxies' physical properties of interest. 

\begin{figure*}[htb!]
\centering
\includegraphics[width=0.9\textwidth]{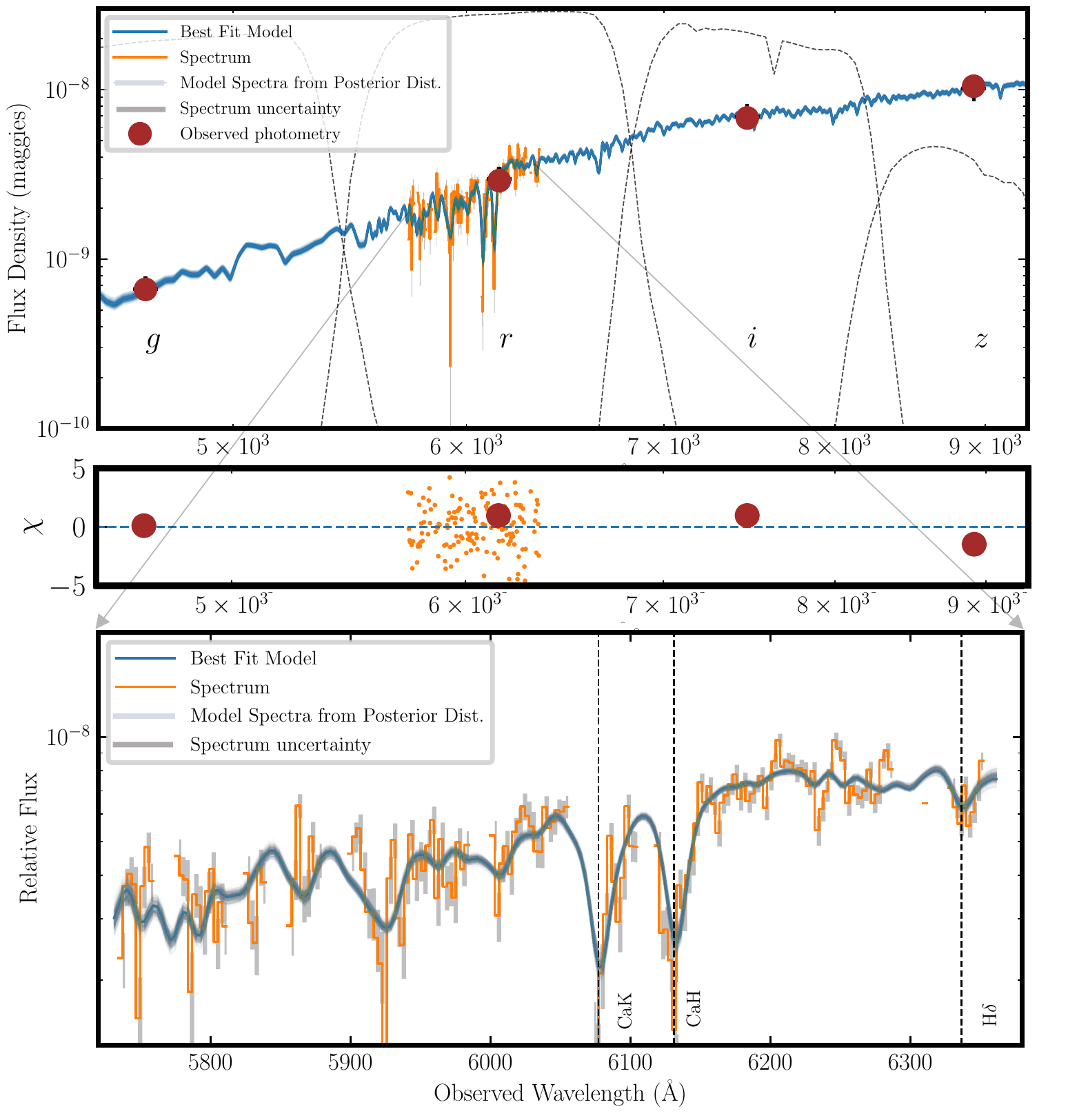}
\caption{(Top) SED of a massive quiescent member galaxy of SPT-CL J2335-4544 at $z=0.55$, shown as flux density (maggies, Jy/3631) vs. observed wavelength (\AA). Model fits (blue) to photometry (brown) and spectrum (orange) are shown. (Middle) Residual ($\chi$) values for spectrum and photometry. Photometric data considered here is precise (uncertainties $< 0.03$ mag). (Bottom) Zoomed-in version of the spectrum, uncertainty and best-fit model.}
\label{fig:fit_sgal}
\end{figure*}

We bin galaxies along the following axes to generate subsamples for stacking:

\begin{itemize}
    \item \textbf{Galaxy Stellar Mass (\Mstar)}: as calculated by SED fitting for a given SFH model.
    \item \textbf{Observed Redshift}: redshift measured from spectroscopy (see B16 and K19).
    \item \textbf{Final Cluster descendant mass, log(M$_{500c,SZ, z=0}/\Msolar$)}: as classified by simulation-based predictions \citep{Fakhouri2010, mcdonald2017} to determine nominal evolutionary paths for SPT clusters across redshift. We sort clusters based on whether their final cluster mass at redshift 0 (M$_{500c,z=0}$) would fall above or below the locus for log(M$_{500c,SZ, z=0}/\Msolar$)=15 — a value of convenience chosen because it splits the cluster sample into two approximately equal parts. \GK{In the B16 cluster sample, we anticipate a uniform distribution of galaxies as a function of clustercentric radius and redshift and hence do not further divide the galaxy sample as a function of radius (the cluster size does not change significantly across the redshift range 0.3$<$z$<$0.9, and the spectroscopic slit sizes are much smaller than cluster radii at a given redshift; see \cite{muzzin2012}}.
    
\begin{figure}[htb!]
\hspace{-0.5cm}%
\includegraphics[width=0.52\textwidth]{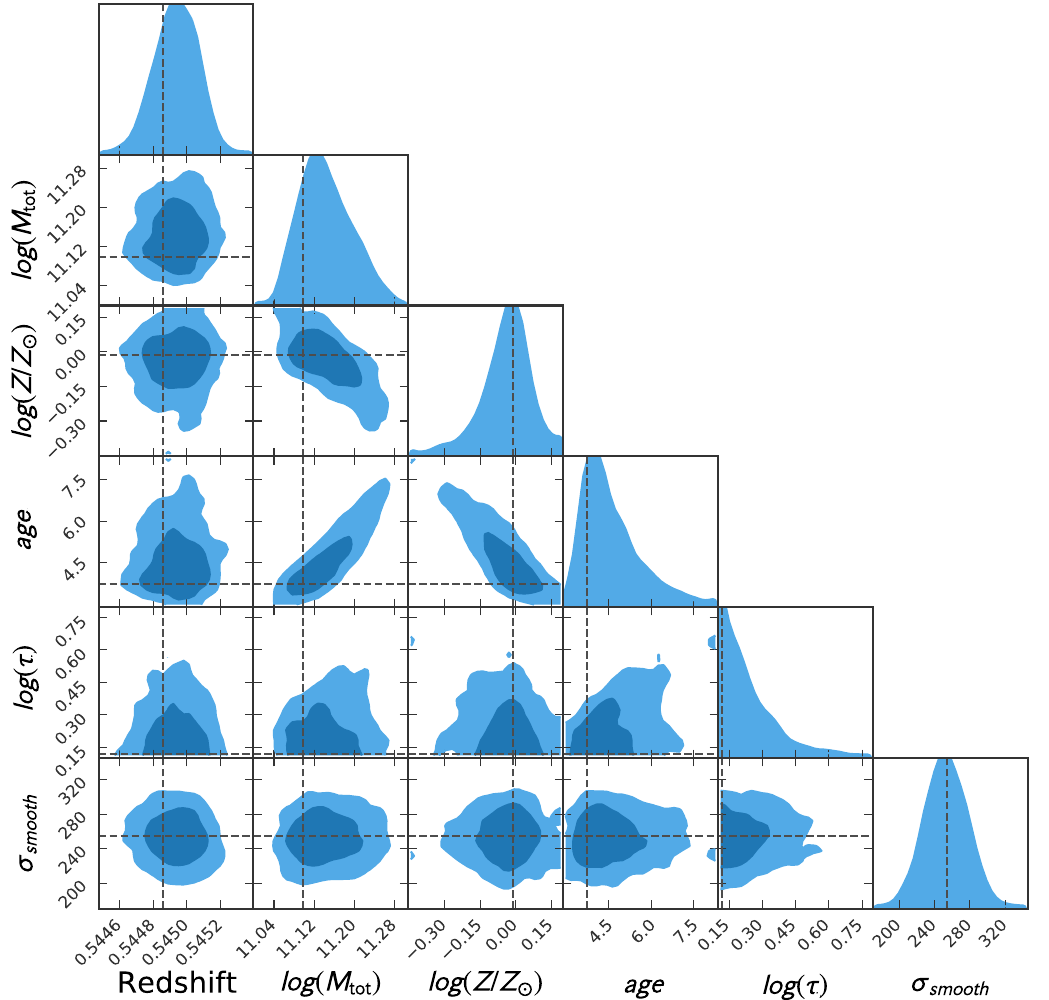}
\caption{Corner plot with posterior distributions and correlations for inferred parameters in the \code{Prospector} SED fitting analysis for a single member galaxy of SPT-CL J2335-4544 at $z=0.55$.}
\label{fig:corner_sgal}
\end{figure}

    \item \textbf{Phase-space location}: \GK{It is common to consider cluster member galaxy properties mapped to cluster-centric radius (e.g., scaled by r$_{500c}$, r$_{200c}$, or virial radius R$_{virial}$; \citealt{ellingson2001,wetzel2012,brodwin2013,strazzullo2019}), and stack observations across a wide range of cluster mass for fixed radius. This has value because there will be some degree of sorting of galaxies by their accretion history i.e., the galaxies first added to the building cluster will tend to appear closer to the cluster center and with more recent additions often observed further out in projection. In this paper, we choose to use a more direct proxy for accretion history, namely the infall time p = r$_{projected}/r_{500c}\times v_{peculiar}/\sigma_{v}$ \citep{noble13,pasquali2019}. Much like the scaling for R$_{200}$, the scaling for cluster core size and velocity dispersion captures the relationship between observed position of the galaxy in phase-space and the cluster virial radius. We use M$_{500c,SZ}$ to compute r$_{500c}$ with cluster SZ centroids, and use dispersion $\sigma_{v}$ values as calculated in B16 and B17. For the spectroscopic sample of SPT cluster members in B16, Kim et al. (in prep) have determined a value of $p<0.4$ implies early or mixed infall, and $p>0.4$ implies late infall of the galaxy into the cluster's potential well. We use these cuts to distinguish galaxy subpopulations by their accretion history.}
    
    \end{itemize}

We determine bin boundaries using a number of factors, e.g., having similar number counts in each stack bin (for galaxies, as well as clusters), physical considerations (e.g., dynamical infall timescales of galaxies in clusters to match redshift bin sizes;$\sim$ 1-1.5 Gyr), having $>30$ galaxies per bin for a given subsample (except in the highest redshift bin). See a summary of bin criteria and description in Table 2.

To produce a median-stacked spectrum from all of the spectra that contribute to a chosen bin, we proceed as follows: First, each spectrum is normalized by the median flux value between rest-frame 4020-4080\AA, a wavelength region lacking any strong spectral features. In this normalization, as in all other aspects of the stacking algorithm, we consider only the spectral pixels not previously masked due to uncertain sky line subtraction. To further guard against outlier pixel values across the sample that weren't previously masked, the stacking algorithm exclusively uses median rather than mean values. At a given pixel in the stacked output spectrum, the algorithm considers as inputs all unmasked spectral pixels that contribute at that wavelength, each of which has been renormalized and resampled as above. To capture the uncertainties on each input flux, while using median estimates exclusively, the algorithm then calculates 10000 instances of the median stack by Monte-Carlo sampling the input flux values with their uncertainties, and taking the median value of each instance. We take the median (50th percentile) of these 10000 instances as the stacked flux. To calculate uncertainties, we bootstrap the above process and use the standard deviation of the resulting distribution of stacked flux values at a given pixel as the uncertainty on the above flux value. Further details of these uncertainty calculations compared against other methods are given in Appendix B.

\section{Results}
\label{sec:results}

\subsection{SED Fitting}

In Figure \ref{fig:fit_sgal}, we show an example of observed photometry, optical spectrum, and the best-fit SED models for a single member galaxy in the cluster SPT-CL J2355-4544 at $z=0.545$, fit via Model A. Figure \ref{fig:corner_sgal} shows a corner plot with posterior distributions of the various fit parameters. We find the best fit total stellar mass formed to be $M_{total}$=$1.32^{+0.11}_{-0.10}\times10^{11}$ \Msolar, best-fit metallicity to be log(Z/Z$_{\odot}$) = -0.03$^{+0.07}_{-0.10}$, and the best-fit dispersion $\sigma=252^{+24}_{-21}$kms$^{-1}$ (instrumental dispersion convolved with intrinsic velocity dispersion). Under the assumption of a delayed-$\tau$ SFH, the best-fit age=$4.35^{+1.25}_{-0.62}$ Gyr and $\tau=0.20^{+0.13}_{-0.08}$ Gyr, making this a galaxy that formed a majority ($>50\%$) of its stars rapidly at $z>1.5$.

\begin{figure}[htb!]
\hspace{-0.5cm}%
\includegraphics[width=0.52\textwidth]{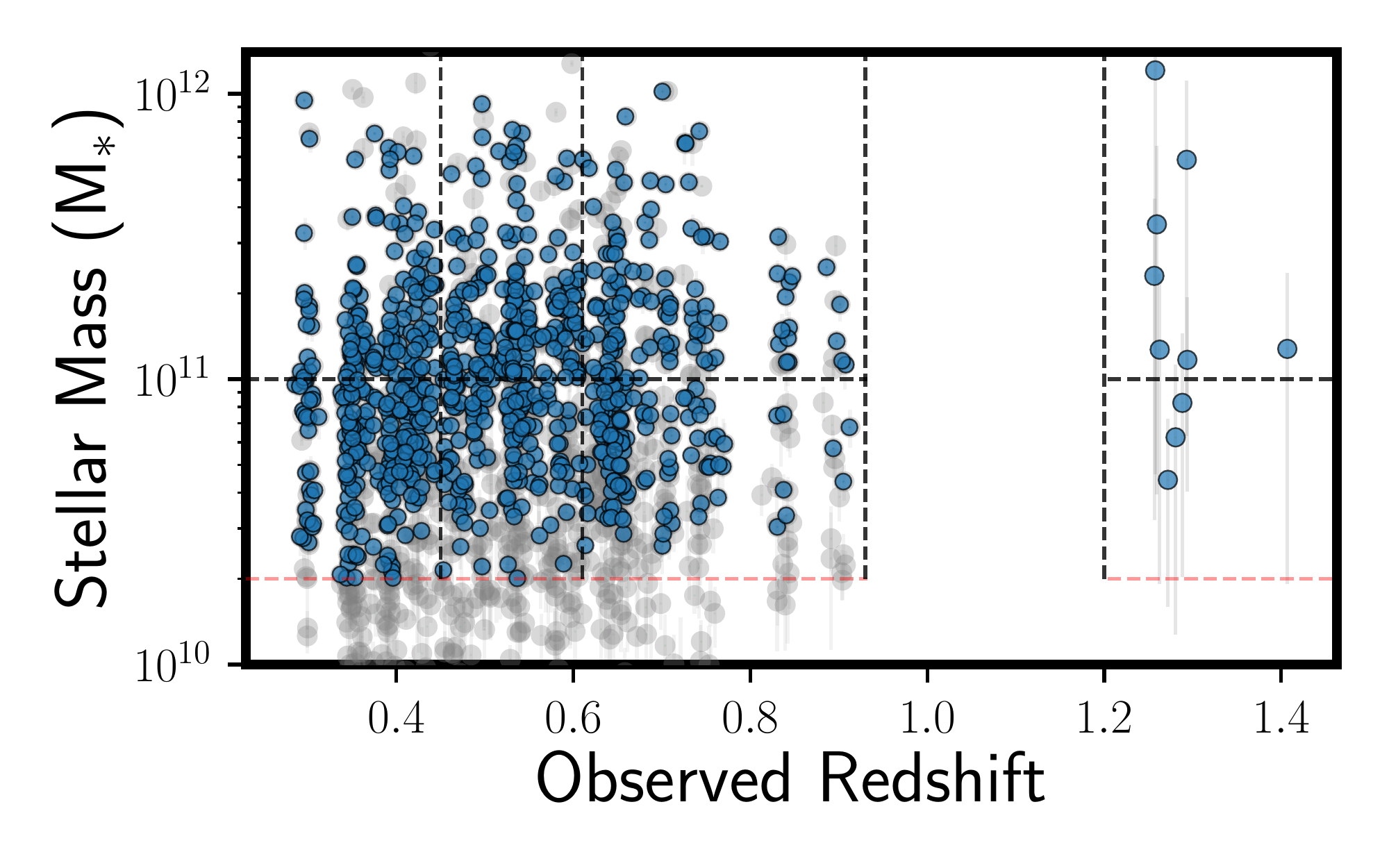}
\caption{Stellar mass (\Mstar) as a function of redshift for member galaxies in the SPT-GMOS survey for clusters at 0.3 $< z <$ 0.9, as characterized by an SED fit to individual galaxies via a delayed-tau SFH model (i.e. Model A, grey points). We exclude a small fraction of galaxies with masses $< 2\times10^{10} M_{\odot}$ (logM$<10.3$) to create a uniform lower limit on the galaxy masses and median mass per bin; the quiescent galaxies considered in this sample are marked with blue points. The dotted lines mark the stellar mass and redshift bins used in this work.}
\label{fig:mstar_dtau_met}
\end{figure}

\subsection{Stellar masses}

Our fitting framework calculates total stellar mass formed in the duration of each galaxy's SFH (M$_{total}$, in units of \Msolar). \code{Prospector} allows us to model the remnant stellar mass (the parameter of interest) for each galaxy, accounting for 20-40\% mass loss from winds and supernovae for a given SFH model. Throughout this work, we refer to the `remnant stellar mass' as stellar mass, unless otherwise noted (M, in units of \Msolar, or log(M/\Msolar)). 

The median stellar mass for the fiducial model across the sample is logM = 10.90, with a range of 10.3$<$logM$<$12.0 (see Figure \ref{fig:mstar_dtau_met}). The signal-to-noise ratio cut (SNR$>5$) implemented here, in combination with the stellar mass cut, results in a cluster galaxy sample with a uniform stellar-mass distribution (flat lower-mass limit and similar median stellar mass per bin) as a function of redshift (for a discussion on SNR, see Appendix D). We also note that we keep consistent the rest-frame optical spectral features that allow us to measure ages and metallicities uniformly (as noted in previous sections), while the photometry for each galaxy — which is a dominant contributor to the calculation of stellar mass M — samples different portions of a given galaxy's SEDs. 

We note that for calculating stellar masses in galaxies, parametric SFHs such as a delayed-tau model show a difference of as much as 0.1-0.2 dex when compared with stellar masses calculated via non-parametric SFHs \citep{carnall2019,leja2019,leja2019b,lower2020}, though this difference is much more prominent in samples of star forming galaxies compared with quiescent galaxies. We note this potential systematic in stellar mass, when comparing results in this work with inferences in the literature.

\begin{deluxetable}{cc}
\label{tab:definitions}
\tablecolumns{2}
\tablewidth{0pt}
\tabletypesize{\scriptsize}
\tablecaption{\bincaption}
\tablecaption{Definitions of age metrics used in this work}
\tablehead{
\colhead{Parameter}&\colhead{Description}\\
}
\startdata
& Mass-weighted age; lookback time\\
$t_{50}$ &  from the redshift of observation when \\
& 50\%of a galaxy's total stellar mass \\
& M$_{total}$ was formed\\
\hline
& Age of the Universe when a galaxy has \\
Age of Universe ($t_{50}$) & formed  50\% of its total stellar mass \\
& (for the assumed cosmology)\\
\hline
$z$ (Age of Universe at $t_{50})$ & Formation Redshift; redshift \\
& at Age of Universe ($t_{50}$)\\
\enddata
\end{deluxetable}


\subsection{Star Formation Histories of Individual Galaxies}\label{sec:sfh_gal}

Using the delayed-tau SFH model, we constrain the age and e-folding time ($\tau$) of each quiescent cluster galaxy in our sample. One of the biggest advantages of a functional form of SFH for a given galaxy is the ability to physically interpret the different stages of galaxy evolution i.e. a nominal star formation start time, a peak of star formation activity, and declining and subsequently quiescent evolution. To consolidate the two parameters age and $\tau$ into one physically interpretable parameter, and to facilitate comparison with studies of massive galaxies exploring mass assembly, we do the following:

\begin{enumerate}

\item We calculate the integral of the assumed delayed-tau SFH (see Equation 1). The normalization of the integral corresponds to the total mass formed $M_{tot}$ of the galaxy, a parameter being fit in the SED fitting process. 
\item For quiescent galaxies in our sample, we define t$_{50}$ as lookback time from the redshift of observation to when the galaxy has formed 50\% of its total stellar mass, or its mass-weighted age. We acknowledge that many studies also use t$_{30}$, t$_{70}$ and t$_{90}$ as parameters of similar interest (e.g., \citealt{pacifici2016}). The definition of t$_{50}$ we use in this work is similar in nature to the mean stellar age or mass-weighted age for a delayed-tau SFH. This allows us to compare our results with mass-weighted age calculations for massive quiescent galaxies in the literature. 
\end{enumerate}

The age of the Universe at the median t$_{50}$ across the sample is $2.3 \pm 0.3$ Gyr, corresponding to a formation redshift z$(t_{50})=2.8\pm0.5$ (see Section \ref{sec:zform_mstar} for more details).

\begin{figure*}[htb!]
\centering
\includegraphics[width=0.96\textwidth]{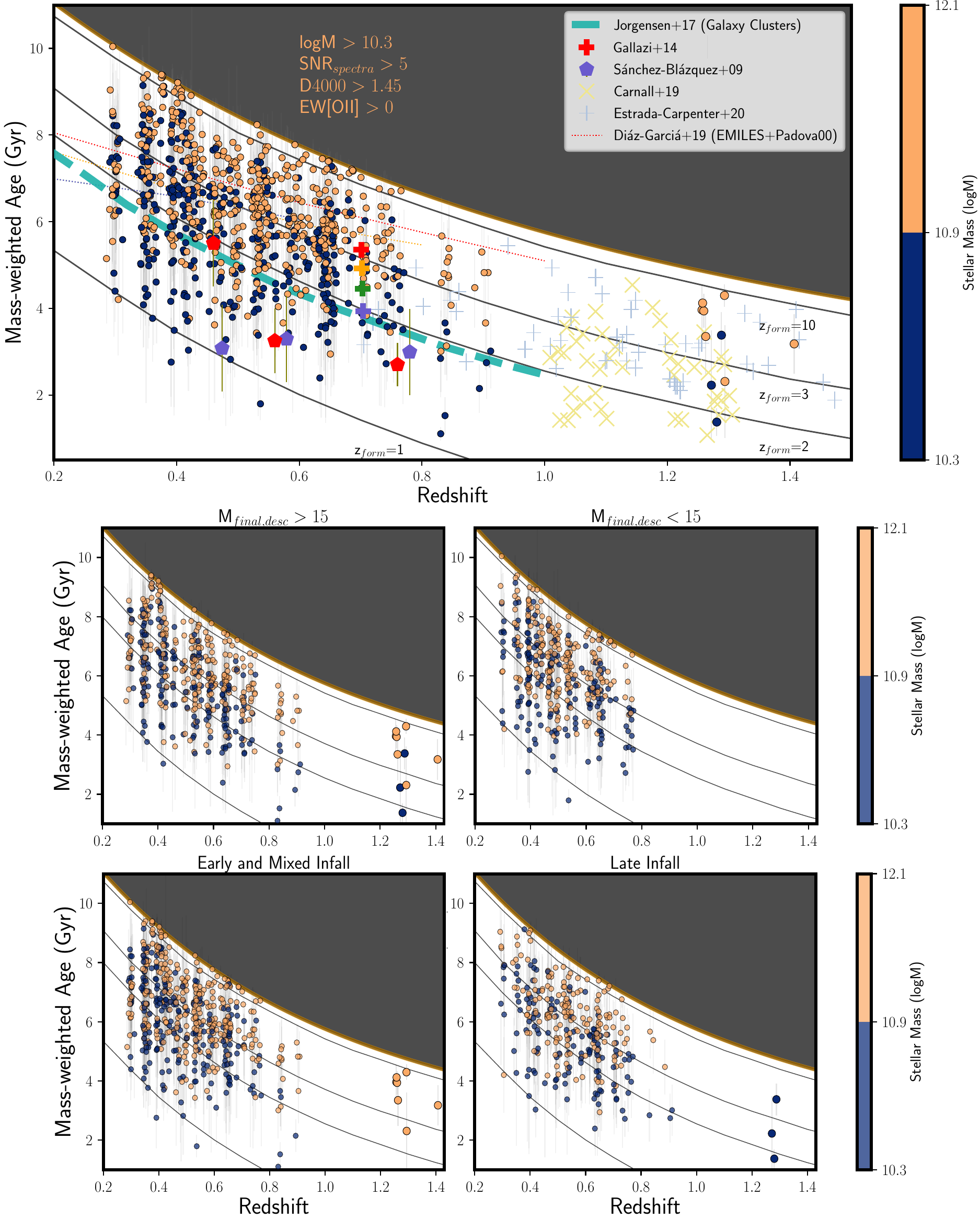}
\caption{(Top) Mass-weighted Age (Gyr) as a function of observed redshift. The circle points represent ages of galaxies from this work, with blue (orange) points representing galaxies with low (high) stellar mass, as calculated via the SED fitting analysis. These measurements are compared to a wide range of literature on cluster galaxies and massive quenched galaxies in the field (see Section \ref{sec:ages}) e.g., \cite{sanchezblazquez2009} and \cite{gallazzi2014} measure ages for quiescent galaxies at 10$<$logM$<10.4$ (purple) and logM$>11.2$ (red) (see Table \ref{table:papers}). (Middle) Age vs redshift for galaxies in clusters with M$_{final,desc}>$15 (left) and M$_{final,desc}<$15 (right). (Bottom) Age vs redshift for galaxies with early and mixed infall (left) and late infall (right), as measured from their phase-space location in velocity-radius space. Black lines indicate the age of simple stellar populations (SSPs) with different formation redshifts as labeled in the top panel.}
\label{fig:age_z_all}
\end{figure*}
\begin{figure*}[htb!]
\centering
\includegraphics[width=0.9\textwidth]{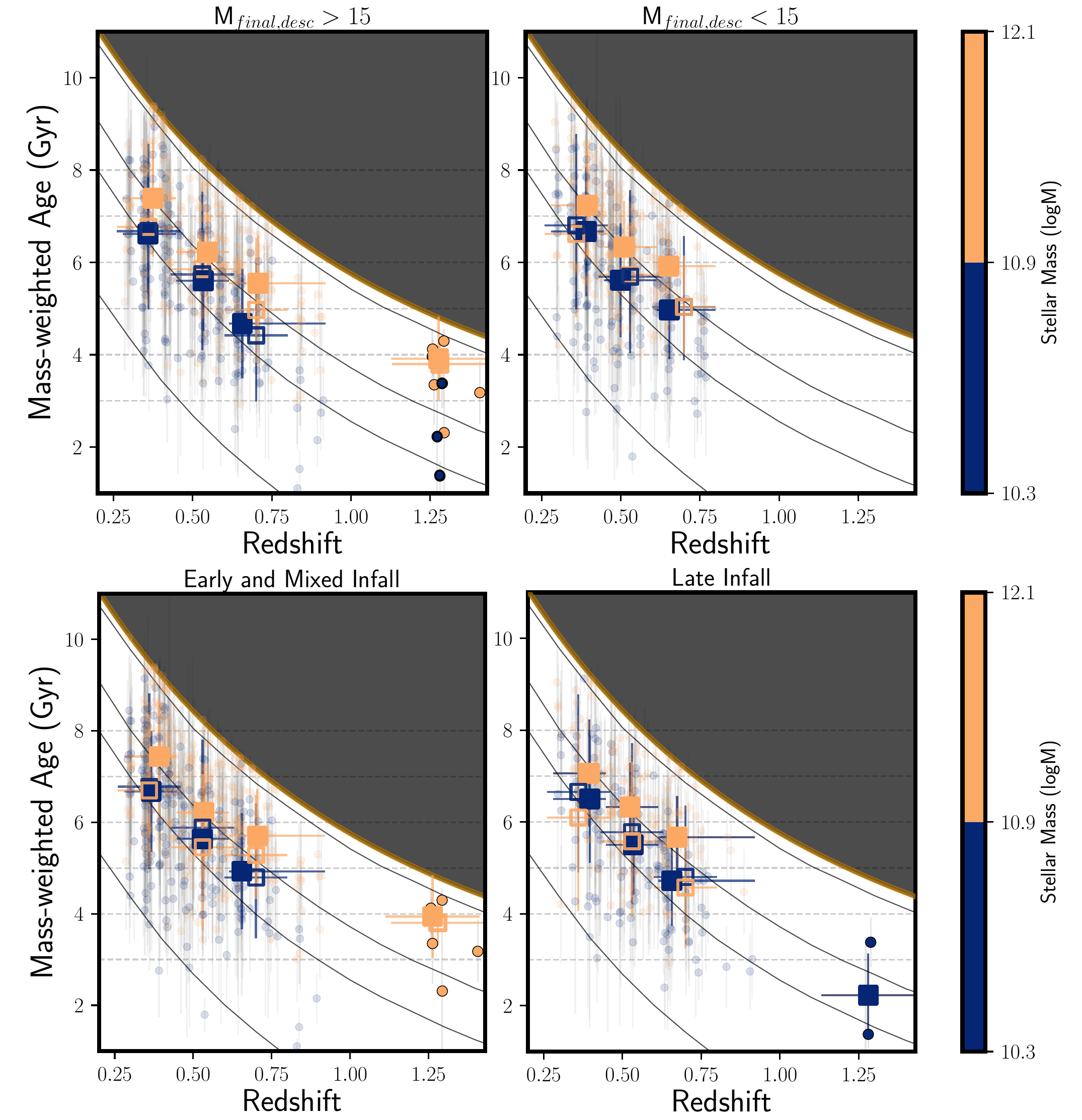}
\caption{(Top) Mass-weighted Age (Gyr) as a function of observed redshift (same as Figure \ref{fig:age_z_all}), to illustrate aggregate and stacked spectra properties of each subpopulation considered in this work. Solid filled squares correspond to median ages per redshift and stellar mass bin (with standard deviation of the ages represented by the error bars). Non-filled squares correspond to stacked spectra ages.}
\label{fig:analysis_and_stack}
\end{figure*}

\section{Ages of Stellar Populations in Cluster Quiescent Galaxies} \label{sec:ages}

The objective of this study is to constrain formation redshifts and stellar masses in massive quiescent galaxies in galaxy clusters. and address the dependence of these properties on accretion history of the galaxies and the cluster mass assembly pathways. 

In particular, we characterise these variables as:\\
a) The final descendent galaxy cluster mass log(M$_{500c,SZ, z=0}/\Msolar$), a proxy for the mass evolution of the host galaxy cluster, and \\
b) The nominal infall time of the galaxy within the cluster, given by a galaxy's phase space location.

Here, we discuss the ages and formation redshifts of these galaxies, and compare these to other massive and quiescent cluster and field galaxy studies. In this section, we refer to mass-weighted ages (t$_{50}$) as \textbf{ages}, and refer to the epoch corresponding to that age as \textbf{age of the Universe at (t$_{50}$)}, unless otherwise specified. See \tabref{definitions} for a summary of key age and redshift metrics used in this work.

\subsection{Mass-Weighted Ages vs Redshift}

\subsubsection{Low-z Galaxies} \label{sec:age_lowz}

\newcommand{\arccaptionee}{Description of studies of massive quiescent galaxies used in this work for comparison}
\newcommand{\arccommentsee}{SSP = single stellar populations, CSP = composite stellar populations.  $^1$\cite{sanchezblazquez2009} use velocity dispersion $\sigma$ as a stellar mass proxy.}
\begin{deluxetable*}{cccccc}
\tablecolumns{6}
\tablewidth{0pt}
\tablecaption{\arccaptionee}
\tablehead{\vspace{-0.2cm} \\ References & Type & Number of Galaxies & Redshift & Stellar Mass, log(M/\Msolar) & Age Measurement}
\startdata
\cite{estradacarpenter2020} & Field & 100 & 0.7$<z<$2.5 & logM $>$10 & CSP + non-parametric SFH\\
\cite{tacchella2021} & Field & 161 & 0.4$<z<$1.25 & 10$<$ logM $<$12 & CSP + non-parametric SFH\\
\cite{carnall2019c} & Field & 75 & 1.0$<z<$1.3 & logM $>$10.3 & CSP + parametric SFH\\
\cite{diazgarcia2019} & Field & 8500 & 0.1$<z<$1.1 & 10$<$ logM $<$11.2 & SSP\\
\cite{gallazzi2014} & Field & 33 & $z\sim0.7$ & logM $>$10.5 & SSP\\
\cite{jorgensen2017} & Cluster & 221 & 0.2$<z<$0.9 & logM $>$10.3 & SSP\\
\cite{sanchezblazquez2009} & Cluster & 215 & 0.4$<z<$0.8 & $\sigma^1>$100 & SSP\\
\cite{webb2020} & Cluster & 331 & 1$<z<$1.5 & 10$<$ logM $<$11.6 & CSP + non-parametric SFH\\
\hline
This work & Cluster & 837 & 0.3$<z<$1.4 & 10.3$<$ logM $<$12.1 & CSP + parametric SFH\\
\enddata
{\footnotesize \tablecomments{\arccommentsee}}
\label{table:papers}
\end{deluxetable*}

In Figure \ref{fig:age_z_all}, we plot t$_{50}$ (or mass-weighted ages) as a function of galaxy redshift, and compare these to sample ages in other published works on massive quiescent galaxies in the field and cluster environments. We show (with black solid lines) evolutionary tracks of simple stellar populations (SSPs) corresponding to an instantaneous episode of star formation at formation redshifts of $z=$ 10, 3, 2 and 1 to visually assess typical formation redshift ranges for these galaxies (\citealt{tacchella2021}). The colors correspond to remnant stellar mass logM of a given galaxy, divided into two bins.
\GK{Consistent with other studies of large samples of massive quiescent galaxies, we identify a diversity of SFHs across redshift and masses in our sample (16th, 50th and 84th percentile ages as 6.23$_{-1.38}^{+1.41}$ Gyr, and a median uncertainty of 1.22 Gyr)}. We note that the most massive galaxies (dark orange circles in Figure \ref{fig:age_z_all}) are seen to have the largest ages possible at the redshift of observation allowed in our SED models (bound by the age of the Universe) Lower mass galaxies (blue circles in Figure \ref{fig:age_z_all}) prefer a mass-weighted age corresponding to an SSP formation redshift of $z<2$, while the highest mass galaxies show formation redshifts of $z>3$, up to $z>10$. This is consistent with downsizing trends seen in other studies — star formation rate and stellar mass assembly of massive galaxies peaked at earlier times relative to lower mass systems \citep{cowie1996, cimatti2006}, which implies that massive galaxies should form their stellar mass earlier than lower mass galaxies. 

Table \ref{table:papers} summarises the literature we have used extensively in this work for comparing ages and formation redshifts; these are studies of cluster and field galaxy samples across a wide range of stellar masses and redshifts. Also shown are the sizes of the sample, and the methodology used to measure ages — either SSPs, single stellar populations (SSPs), or composites of SSPs (composite stellar populations, CSPs).

In Figure \ref{fig:age_z_all}, we also show results from \cite{estradacarpenter2020} (blue plus points) and \cite{carnall2019c} (yellow crosses) at $z>0.8$, who calculated ages of massive quiescent field galaxies from CSP-based SED models. We plot data from stellar mass bins given in \cite{gallazzi2014} (plus points) and \cite{diazgarcia2019} (dotted lines) at 0.4 $ < z < $ 0.8 where, the mass bins are defined as 10$<$logM$<$10.4 (purple), 10.4$<$logM$<$10.8 (yellow), 10.8$<$logM$<$11.2 (green) and logM$>11.2$ (red).
These studies show ages calculated via SSP-based models (which tend to be lower, and bias ages towards the most recent episode of star formation; see e.g., \citealt{carnall2018}). 

There are limited cluster-based studies that calculate galaxy properties using SED models that are non-SSP based (i.e. without assuming an instantaneous burst of star-formation) in this redshift range. \cite{jorgensen2017} (cyan dashed line) and \cite{sanchezblazquez2009} (green pentagon points) use SSP-based models to calculate ages of galaxies from clusters between $0.2 < z < 0.9$ (with a sample lower limit of masses and velocity dispersions of galaxies similar to this work). While ages from these lowest mass galaxies are consistent with these studies, we anticipate a systematic bias of $>1$ Gyr in these studies given model assumptions \citep{carnall2019}, when compared to this work.

In middle panels of Figure \ref{fig:age_z_all}, we plot a subset of galaxies divided by membership in clusters above or below log(M$_{500c,SZ, z=0}/\Msolar$) = 15. Comparing the two subsets, we do not see a substantial difference in ages and stellar mass distribution as a function of redshift (except galaxies in the redshift $z>0.61$ clusters). This is explored further in Section \ref{sec:median_stack} and Figure \ref{fig:analysis_and_stack}.

Bottom panels of Figure \ref{fig:age_z_all} show the subset of galaxies tagged as early+mixed infall times (bottom left) and late infall times (bottom right). \GK{We note that the oldest galaxies in our sample (at z$_{form,SSP}>10$) are mostly located in the early+mixed infall subsample — 28 galaxies (26 with high stellar mass (orange)), relative to only 7 in the late infall subsample. This indicates that these galaxies have spent one or multiple turnaround times around the center of a cluster gravitational potential well and have the highest mass-weighted stellar ages,} consistent with a hierarchical picture of a grow-and-quench evolution mechanism; see Section 1 of \cite{tacchella2021} and references therein.

\subsubsection{Mass-weighted ages of galaxies observed at $z>1.2$}

The 10 highest redshift massive quiescent galaxies in our sample span 1.22 $ < z < $1.42, and belong to the high log(M$_{500c,SZ, z=0}/\Msolar$)$>15$ bin. The median and 16th-84th percentile range of ages is t$_{50}$ = 3.4$_{-1.1}^{+0.7}$ Gyr. Three of the galaxies have a phase-space location corresponding to late-infall time (median age t$_{50}$ = 2.2 Gyr), while 7 galaxies belong to the early+mixed infall bin (median age t$_{50}$ = 4.1 Gyr); see middle and bottom panels of Figure \ref{fig:age_z_all}. 

The range in mass-weighted ages is consistent with the diversity seen in massive field galaxy samples in \cite{carnall2019c} (2.7$_{-1.1}^{+0.8}$ Gyr) and  \cite{estradacarpenter2020} (3.0$_{-0.8}^{+0.9}$ Gyr), studies of massive quiescent galaxies sampling similar redshift ranges. The lack of a substantial offset in median ages of cluster members and standard deviation of ages compared with field galaxies (albeit for 10 galaxies only) is consistent with other cluster and field galaxy studies \citep{raichoor2011,webb2020}. Moreover, qualitatively, if we consider galaxies in the log(M$_{500c,SZ, z=0}/\Msolar$)$>15$ bin, we find that the formation redshifts of high-z galaxies are consistent with those of the most massive low-z galaxies (i.e., these can be considered to be antecedents of low-z galaxies), and consistent with a purely passive evolution scenario.

\begin{figure}[ht!]
\centering
\includegraphics[width=0.5\textwidth]{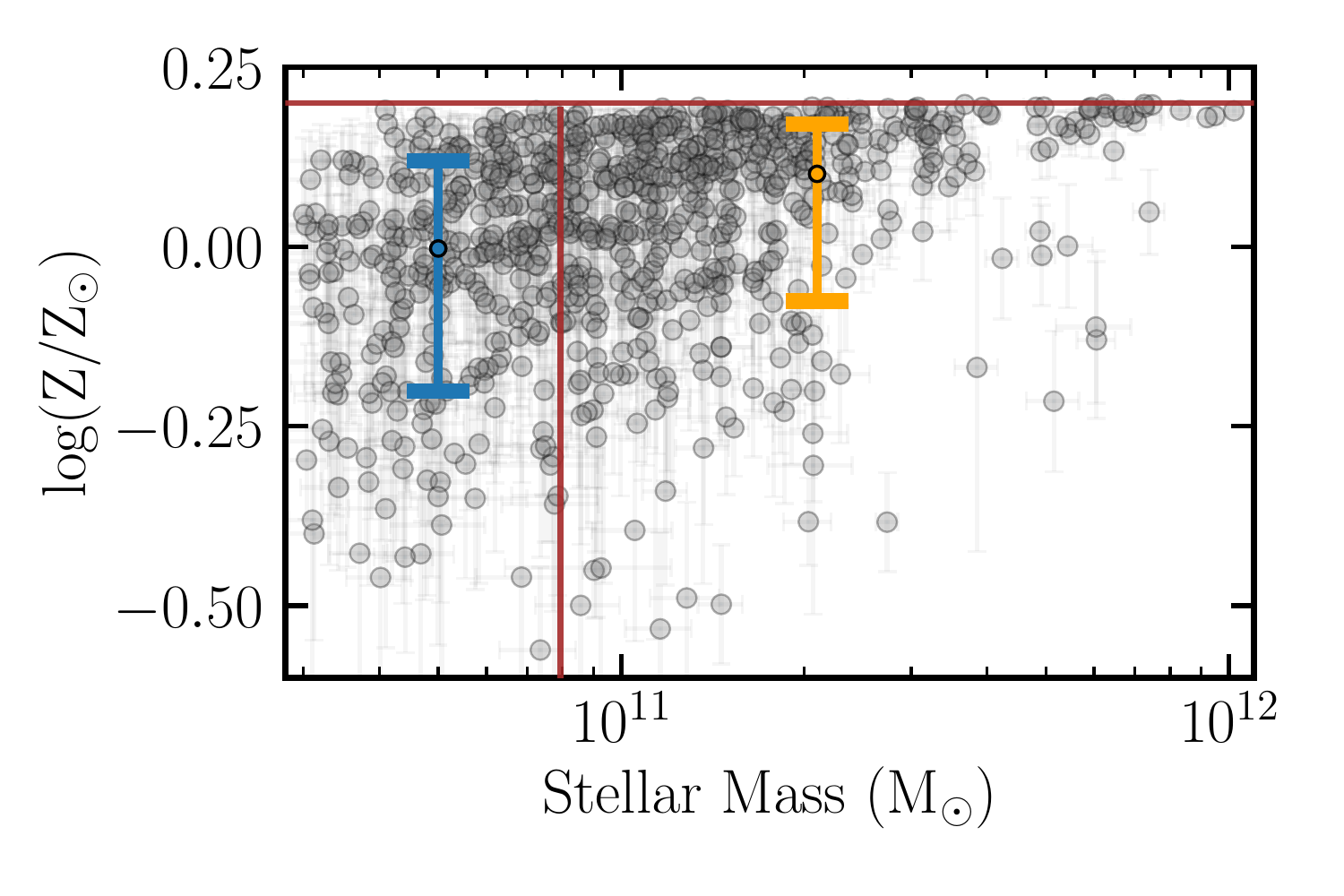}
\caption{Metallicity (log$(Z/Z\odot)$) distribution for galaxies considered in this work, as a function of stellar mass. The red horizontal line corresponds to the prior boundary, and the red vertical line is the bin boundary for stellar mass considered in this work. Median, 16th and 84th percentile distribution for metallicity is shown in blue (low mass) and orange (high mass). The median uncertainty in log$(Z/Z\odot)$ is 0.15 dex.}
\label{fig:met_small}
\end{figure}

\subsubsection{Median and Stack Properties} \label{sec:median_stack}

To demonstrate the aggregate spectral properties of our galaxies, and reliably measure median galaxy properties in our $z>1.2$ sample (with only low SNR spectra that are seriously compromised by sky-subtraction residuals at some wavelengths), we illustrate mass-weighted ages of stacked spectra as a function of redshift in Figure \ref{fig:analysis_and_stack}. We show ages from stacked spectra (with hollow squares) and median ages per redshift and stellar mass bin (with filled squares) for a given subpopulation divided by either cluster mass or phase-space location, akin to discussions in the previous sections. The highest redshift bin ($z>1.2$) only has 3 galaxies in the late infall subpopulation, and is not stacked.

Our stacking results are observed to be consistent with the median properties of galaxies in a given bin — we recover marginal downsizing (increasing formation redshifts with increasing stellar mass), and a marginal increase in formation redshifts with increasing observed redshift in the highest redshift bins (see SSP evolutionary tracks in Figure \ref{fig:analysis_and_stack} corresponding to z$_{form}$=10, 3, 2 and 1) — which gives credence to the spectral properties of the stacked spectrum of the z$>1.2$ stack. We also advocate for the uncertainties in each stacked spectrum to reflect the diversity in mass-weighted ages of the constituent galaxies, which is achieved here by Monte-Carlo sampling individual galaxy spectra and measuring the standard deviation of the sampled spectra. This methodology does not underestimate uncertainties in each stack (unlike the stacking methodology where the uncertainty is measured by obtaining the uncertainty on the mean flux in the Monte-Carlo sampled spectra per bin). 

Additionally, we observe significant impact of metallicity on obtaining consistency between stacked spectra ages and median ages per bin. If the constituent galaxies cover a wide range of metallicities — $>0.1$ dex in log(Z/Z$_{\odot}$) — it is challenging to ascribe appropriate priors to the metallicity parameter in constructing a stacked spectrum. This is especially true for galaxies with stellar masses logM $< 10.90$; see Figure \ref{fig:met_small} that shows the distribution of log(Z/Z$_{\odot}$) as a function of logM, and median and 16th-84th percentile range of metallicities per stellar mass bin. 

Moreover, we note a particularly wide distribution of metallicities in galaxies in the lowest redshift bins ($0.29<z<0.61$; $\sim$ 0.1 dex) relative to higher redshifts ($z>0.61$; $\sim$ 0.05 dex). This is consistent with the idea that at high redshifts, massive quiescent galaxies have a restricted set of pathways to achieve quiescence (D4000 $>1.45$ and a lack of [OII] emission), whereas at low redshifts galaxies can achieve quiescence through multiple pathways (regardless of binning by cluster mass or phase space location). 

A detailed discussion of the metallicity distribution of individual galaxies and the alternate stacking method is given in Appendix A and B, respectively.

\subsection{Formation Redshifts}

\begin{figure*}[htb!]
\centering
\includegraphics[width=1.01\textwidth]{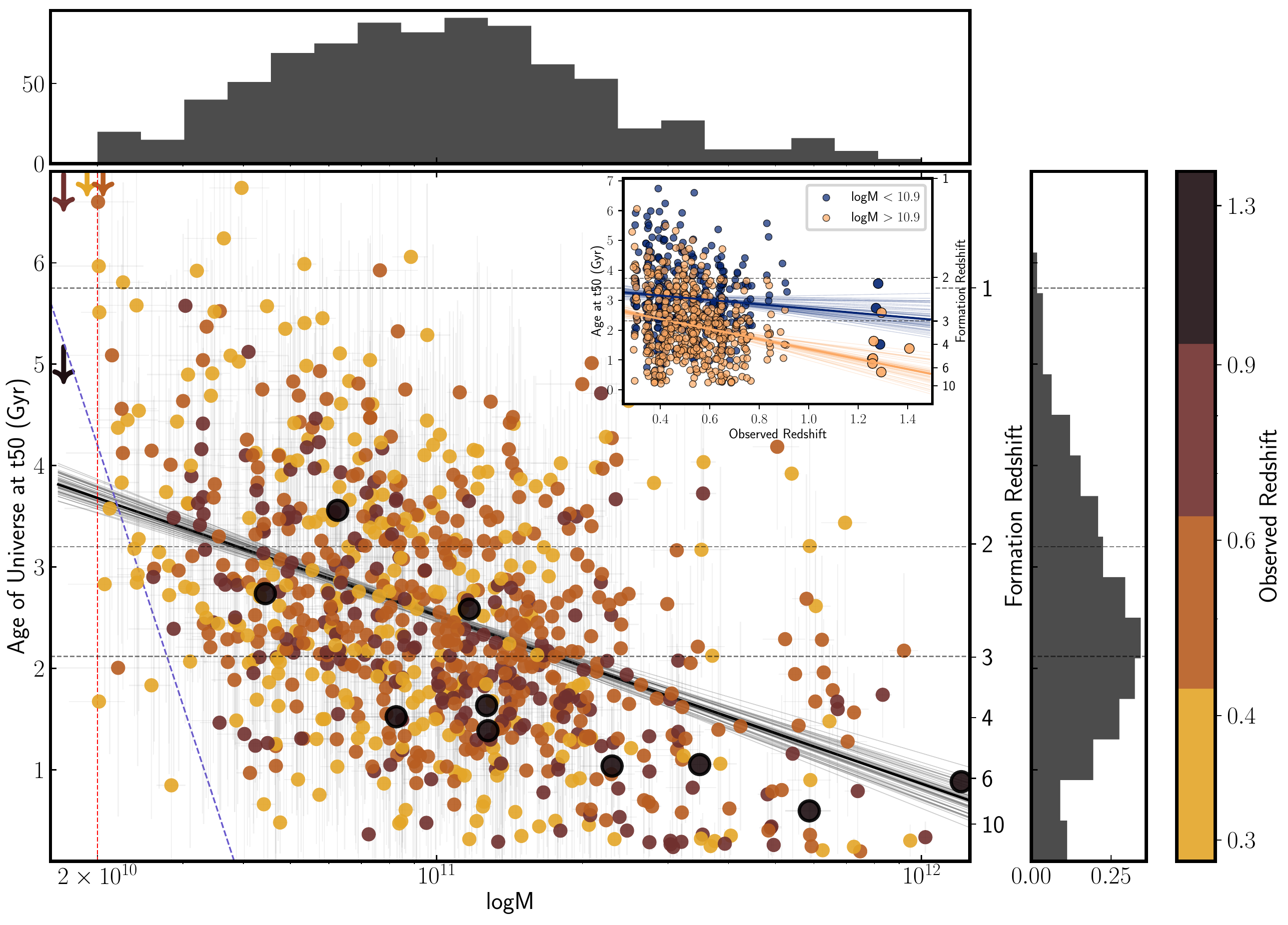}
\caption{Age of Universe at which the galaxy has formed 50\% of its mass (corresponding to t$_{50}$, in Gyr) vs logM for quiescent galaxies considered in this work. Color at each point denotes the observed redshift, with orange points at z$\sim$0.3 and large black points at z$\sim$1.3, the highest redshift galaxies in the sample. Best-fit age-mass relation is plotted in black, with grey lines sampling 50 lines from randomly sampled combinations of slopes and intercepts in the range of best-fit values and 1$\sigma$ uncertainties. Downward arrows correspond to the maximum age of the Universe allowed for the median redshift in each bin — 5.1 Gyr at $z=1.28$, and 9.1 Gyr at $z=0.37$. Overplotted are the fixed mass cut considered in this work (dotted red line). We also show a fixed detectability (or SNR) cut for the lowest mass galaxies near the B16 survey detection limit (in dotted blue), with the slope of the line equal to the best-fit age-(L/M) relation; see Section \ref{sec:incomplete} for details. (Inset) Age of Universe at t$_{50}$ (Gyr) vs Observed redshift, for galaxies studied in this work, split by stellar mass. }
\label{fig:t50_z_mstar}
\end{figure*}

To quantify the suggestive age differences and downsizing observed in various subpopulations of galaxies considered in this work, and to compare galaxy evolution from different epochs of observation on a common timescale, we plot formation redshifts (corresponding to \textbf{age of the Universe at t$_{50}$}) as a function of stellar mass (logM); see similar analysis of downsizing and trends between observed and formation redshift in samples of field quiescent galaxies \citep{carnall2018,carnall2019c,estradacarpenter2020,tacchella2021}. 

Figure \ref{fig:t50_z_mstar} shows the age of the Universe at t$_{50}$ (in Gyr) vs logM for all galaxies considered in our sample, with color denoting the redshift of observation. The distribution of ages demonstrates both downsizing trends (higher formation redshifts for higher mass galaxies) as well as decreasing formation redshifts with decreasing observed redshifts, consistent with both cluster and field galaxy studies mentioned above. 

The distribution of galaxy ages also shows that the majority of quiescent galaxies in our cluster sample have formed 50\% of their stellar mass between z=2-3. Note that while the highest redshift galaxies in our sample (at z$\sim$1.3, large black points) are not a statistically large sample — 10 galaxies with a median uncertainty $\Delta$t$_{50,age\ of\ Universe}$ = 0.56 Gyr — the individual formation redshifts are consistent with a downsizing trend.

We also fit a linear age-mass relation to each subpopulation, depicted with solid lines in the top panel. The linear fits to each subsample are given by the following model:
\begin{equation}
    t_{50,age\ of\ Universe} (Gyr) = \alpha  \log_{10}\left(\frac{M}{10^{11}\Msolar}\right) + \beta
\end{equation}

The formation ages/redshifts of the full sample of massive quiescent cluster galaxies in this work is described by t$_{50,age\ of\ Universe}$ (Gyr) = $2.52 (\pm0.04) - 1.66 (\pm0.11)$ log$_{10}(M/10^{11}\Msolar)$. This relationship becomes marginally steeper in the highest redshift bins. The best-fit slopes $\alpha$ and intercepts $\beta$ for the binned sub-populations are given in Table \ref{table:sed_slopes}. 

\subsubsection{Age-Stellar Mass Relation and Sample Representativeness}\label{sec:incomplete}

\GK{The spectral data for the B16 cluster galaxy sample were taken with the intent of measuring the same mean signal-to-noise ratio per spectrum across redshift for a given absolute magnitude. The sample is not complete (e.g., in that at no absolute magnitude or cluster-centric radius were all cluster galaxies observed, at any redshift). However, it is a representative sample of quiescent member galaxies in SPT galaxy clusters, intended to have a uniform galaxy mass limit across the redshifts considered. As can be seen in Figure \ref{fig:mstar_dtau_met}, this is approximately true, and we have imposed a lower stellar mass threshold of logM$>$10.3 to further ensure this. Within this sample, the median stellar mass as a function of redshift is flat to within uncertainties. In the analysis presented here, we do not measure the bulk properties of clusters (e.g., luminosity function, stellar mass function, blue fraction, etc.) for which a careful accounting of incompleteness would be critical. Instead, the focus of this study is the spectroscopic analysis of quiescent galaxies that are a representative (and not complete) sample of cluster members.}

\begin{figure}[h!]
\centering
\includegraphics[width=0.5\textwidth]{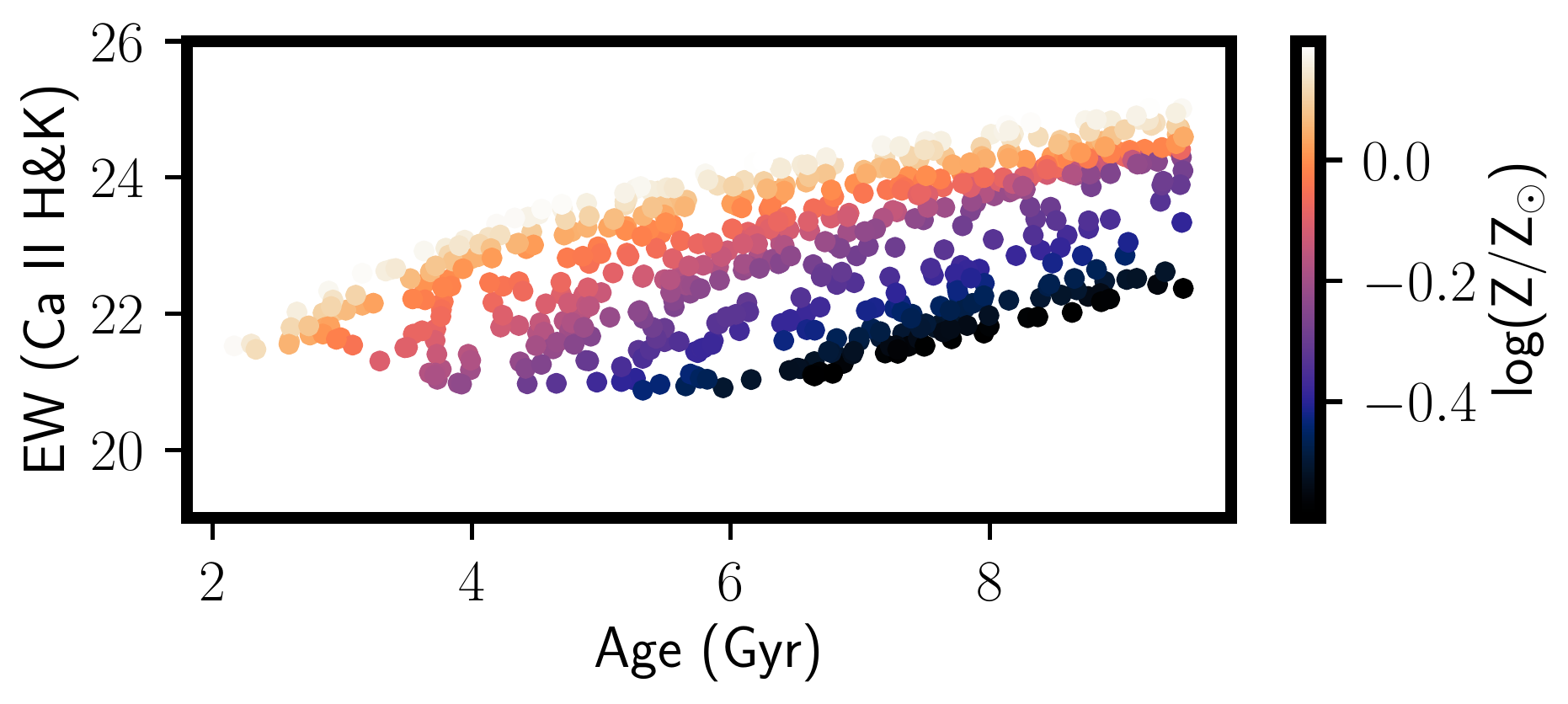}
\caption{Equivalent width of Ca II H$\&$K as a function of age for model SEDs of quiescent galaxies (D4000$>$1.45). Colors correspond to metallicity log($Z/Z_{\odot}$). }
\label{fig:ew_age}
\end{figure}

Nevertheless, because the galaxy spectroscopy discussed here is not a complete sampling of all cluster galaxies within a fixed physical aperture for each cluster, we must also consider whether subtler incompleteness exists that is correlated with any parameter of interest e.g., age, metallicity and stellar mass. Such incompleteness might particularly influence results on the low mass end. For example, if the strengths of primary spectral features used to measure redshifts in B16 have a strong negative correlation with age, we might worry that an absence of old low-mass cluster galaxies is due to a failure to measure redshifts for such galaxies, as opposed to an actual absence of such galaxies in clusters.

We test this by measuring equivalent widths of the features that were primarily used to characterize redshifts in B16 and K19 - Ca II H$\&$K and the G-band. These are the strongest absorption features in quiescent galaxy spectra that are visible across the redshift range considered. We compare these measurements with formation redshifts of galaxies in Figure \ref{fig:t50_z_mstar}, and explore whether we are missing spectroscopic confirmation of galaxies in the bottom left (early formation, low stellar mass) and top right (late formation, high stellar mass) corners. We generate 1000 model SEDs for quiescent galaxies using \code{Prospector} and the models described in Section \ref{sec:sedfit_desc} by randomly sampling parameters from the priors used in our study (Table \ref{table:sed_arc}). For these models, we find that EW(Ca II H$\&$K) and EW(G) in galaxies with D4000$>1.45$ increases marginally with age. Figure \ref{fig:ew_age} shows EW(Ca II H$\&$K) as a function of age for model SEDs. For galaxies with ages $>6$ Gyr (corresponding to a maximum formation redshift of $z=10$ allowed by our sample redshift range), the fractional change in EW(Ca II H$\&$K) ($<10\%$), and the impact on the probability of redshift success is negligible. 

This implies that the probability of redshift success for a given spectrum's signal-to-noise ratio is not negatively correlated with strength of spectral features. We also show that the mean signal-to-noise ratio (SNR) per spectrum does not change as a function of redshift (above a threshold of SNR$>5$; Appendix D, Figure \ref{fig:snr}), as was the intent of B16.

\begin{figure}[h!]
\centering
\includegraphics[width=0.34\textwidth]{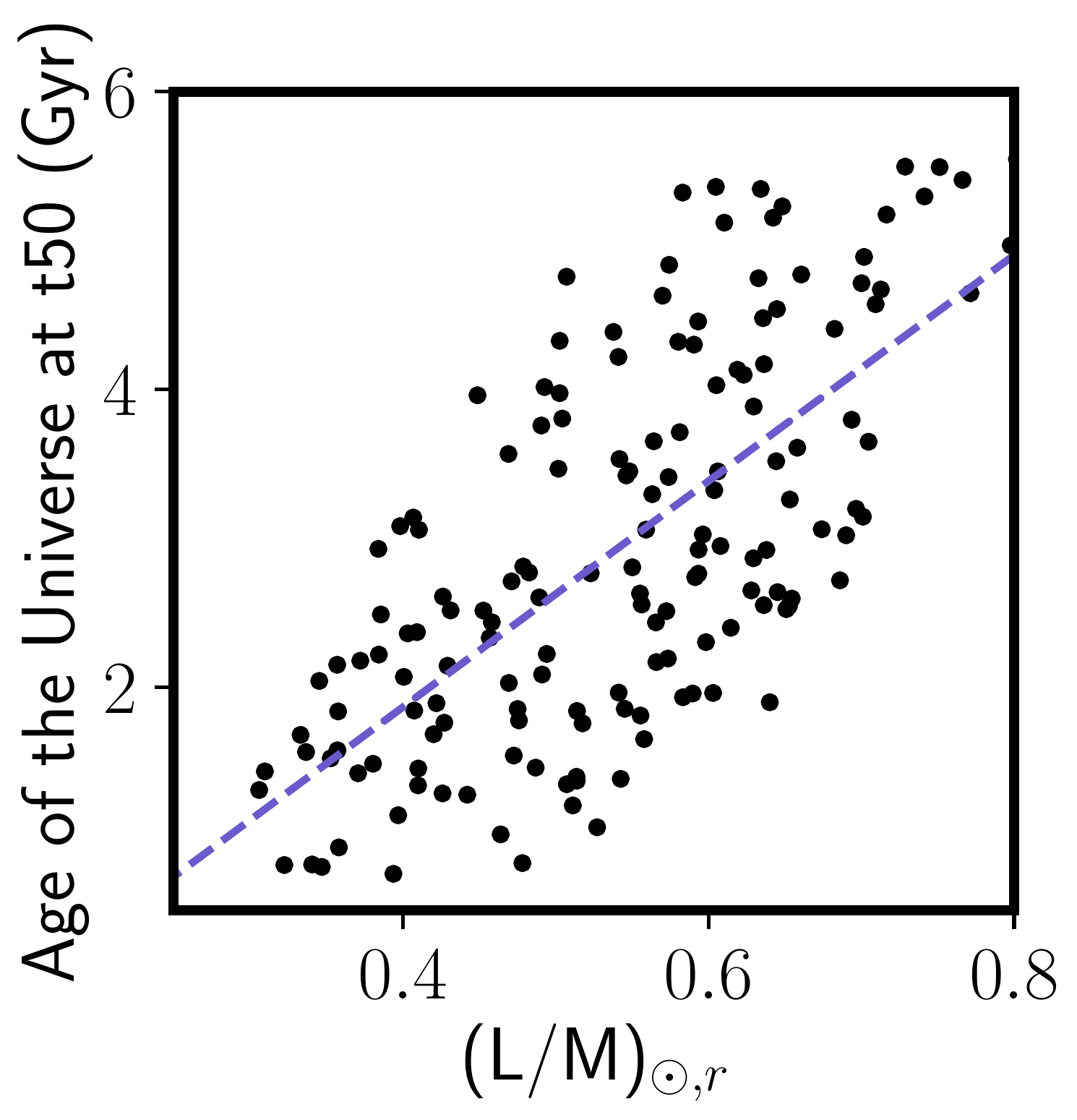}
\caption{Age of Universe at which the galaxy has formed 50\% of its mass (corresponding to t$_{50}$, in Gyr) vs inverse of mass-to-light ratio (L/M) (in solar units) for D4000$>$1.45 quiescent galaxies drawn from model SEDs. The best-fit relation is plotted as a dotted blue line.}
\label{fig:mlratio}
\end{figure}

\GK{While these tests suggest the the B16 input sample has robust sampling in SNR with redshift, and as shown previously has a flat mass cut across redshift, there is a further effect to consider at the low mass end - namely that the mass-to-light (M/L) ratio of a stellar population of fixed mass changes as it ages. To test this further we generate model SEDs for galaxies with a fixed stellar mass (logM=10.4) and fixed redshift (z=0.5), and spanning the same metallicity and age range as galaxies in Figure \ref{fig:t50_z_mstar}. We calculate the formation redshifts and M/L ratios, corresponding to $r$ band (AB) luminosities (a close match to the rest-wavelength interval over which spectra have been fit), in units of (M/L)$_{\odot}$) for D4000$>$1.45 galaxies in this sample. Figure \ref{fig:mlratio} shows the distribution of ages of the Universe at t$_{50}$ (Gyr) as a function of the inverse of (M/L) i.e. light-to-mass ratio; lower values of (L/M) correspond to fainter galaxies for a fixed stellar mass. We also calculate the best-fit age-(L/M) relationship, plotted in blue; this implies that for a fixed stellar mass, our observing strategy preferentially chooses brighter late-formed galaxies and is likely to miss the lowest (L/M) (or low mass early-formed) galaxies. In Figure \ref{fig:t50_z_mstar}, we overplot the flat stellar mass cut (dotted red), and also use the slope from the best fit age-(L/M) (dotted blue) to plot a `fixed detectability' (or equal SNR) line in the lowest-mass end of the distribution. Note that the two lines by choice intercept at the median age (4.2 Gyr) in the lowest stellar mass subsample (logM$<$10.4). Quiescent galaxies in the normal direction to the `fixed detectability' line to the left are less likely to be detected in the spectroscopy sample and those to the right are more likely to be present. While this will slightly shape the distribution in the age of the Universe at t$_{50}$ versus stellar-mass plane, it is not primarily responsible for the observed distribution.}

For completeness we also note that the high-mass late-forming corner of this diagram (i.e., upper right in Figure \ref{fig:t50_z_mstar}) is highly unlikely to be affected by incompleteness that would shape the distribution, as such galaxies would be observed at high SNR. Thus the  upper envelope of the distribution of galaxies in this plane, which traces the same slope as the overall fit, is robust.

We also compare the D4000 spectral index distribution measured in this work to that of other cluster and field galaxy studies \citep{balogh1999,hutchings2000, kauffman2003,noble13,haines2017}. At our cut (D4000 $>1.45$), the D4000 distributions in these studies are similar to this work. We also note that the secondary peak of D$_n$4000 distribution in Figure 21 of \cite{kauffman2003} at D$_n$4000=1.8 differs from the D4000 peak we see in this work (Figure \ref{fig:d4000_o2}) by 0.5$\sigma$.

\subsubsection{Formation Redshift across bins of Observed Redshift}\label{sec:zform_mstar}

To quantify downsizing trends inferred from our SED fitting analysis across our sample, we analyse sub-populations based on the binning criteria described in Section \ref{sec:stack_description}. 

\begin{figure*}[htb!]
\centering
\includegraphics[width=1.0\textwidth]{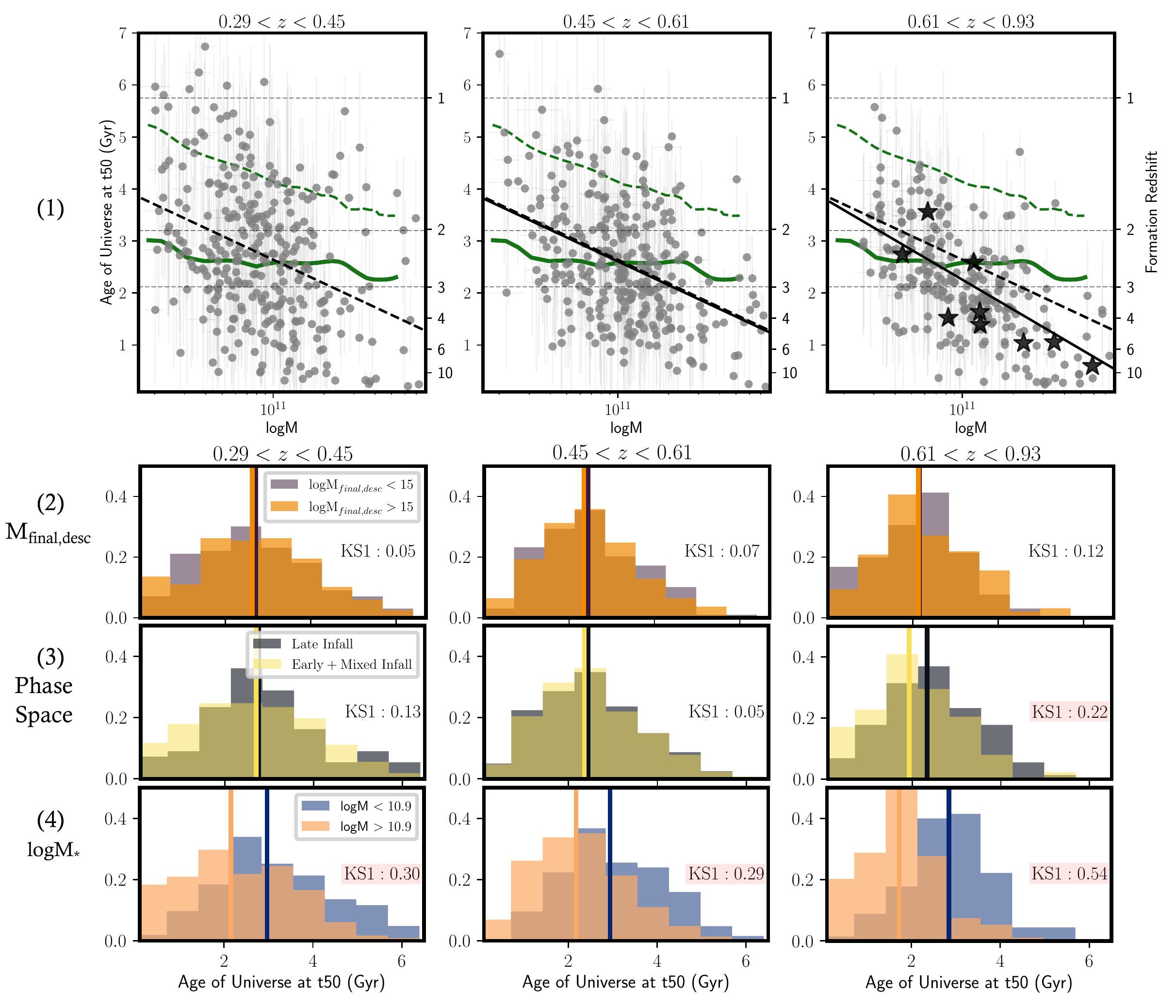}
\caption{(\textbf{Row 1}) Age of the Universe at t$_{50}$ in Gyr (or formation redshift) vs. stellar mass logM for each galaxy. Each subplot displays the galaxy subpopulation in redshift bins $0.29<z<0.45$, $0.45<z<0.61$, and $0.61<z<0.93$ (star symbols in panel 3 correspond to 10 $z>1.2$ galaxies in our sample). Dotted horizontal grey lines represent ages of the Universe corresponding to redshifts 1,2 and 3. Dark green dashed and solid lines represent ages of Universe at formation times vs logM from the TNG100 field galaxy simulations and redshifts at 0.1 and 1 respectively, as seen in Figure 8 of \citealt{carnall2019c}. We fit for a mean relationship between formation redshifts and log(M), indicated by black lines. The diagonal dashed black lines correspond to the mean relationship from panel 1 (the lowest redshift subsample), plotted in all three panels to facilitate comparisons across redshift bins. (\textbf{Row 2}) Distributions of age of the Universe at t$_{50}$ per redshift bin, split by log(M$_{500c,SZ, z=0}/\Msolar$) $>15$ (dark orange) and log(M$_{500c,SZ, z=0}/\Msolar$) $<15$ subpopulations (purple). Vertical lines correspond to median ages. Inset text refers to KS1 statistic values for the pair of distributions in each panel (see Section \ref{sec:ks}). High values of KS1 indicate that the hypothesis that the distributions are the same can be ruled out; statistically significant values are shaded in red. (\textbf{Row 3}) Same as Row 2 panels, but galaxies split by phase space ($r_{projected}/r_{500c} \times v_{peculiar}/\sigma_{v}$), corresponding to early+mixed infall (yellow) and late (black) infall sub-populations. (\textbf{Row 4}) Same as Row 2 panels, but galaxies split by stellar mass, with logM$>10.9$ subpopulation in light orange, and logM$<10.9$ subpopulation in blue.}
\label{fig:t50_logm_alldists}
\end{figure*}

Row 1 of Figure \ref{fig:t50_logm_alldists} illustrates formation redshifts and stellar masses for galaxies divided into three observed redshift bins. On comparing median formation ages of each subpopulation, we see an average downsizing trend of $\sim0.5$ Gyr (across a range of 1.5 dex in logM), which agrees with both field quiescent galaxies at $0.3<z<3$ in \cite{carnall2019} and cluster galaxy work seen at $z>1$ in the GOGREEN survey \citep{webb2020}. The most massive galaxies lie in the bottom right corner of each panel in this plot, as is expected in the scenario of a simple hierarchical structure formation and mass assembly (e.g., \citealt{springel2005}). The diagonal dashed lines plotted in all panels correspond to the best-fit mean age-mass relation for the lowest observed redshift bin ($0.29<z<0.45$), to facilitate visual comparison of the relation gradient across redshift. The highest redshift bin $0.61<z<0.93$ has the steepest gradient, with a $\sim 0.75$ Gyr difference from corresponding galaxies in the lower redshift bins.

Figure \ref{fig:t50_logm_alldists} demonstrates that the two lower observed redshift bins contain galaxies with a diversity of formation redshifts in each stellar mass subpopulation, with similar median formation times and age-mass relations. This indicates that the dependence of formation time on stellar mass does not evolve significantly between $0.3<z<0.6$ ($\sim2$ Gyr in time), over a wide range of stellar masses. For a given stellar mass, this could be either because galaxies in the two lowest observed redshift bins could be derived from the same population, or newer galaxies joining the quiescent population sample the same range of formation redshifts.

In Figure \ref{fig:t50_logm_alldists}, we also overplot age and stellar mass values from two snapshots of the IllustrisTNG simulations at $z=0.1$ (dotted dark green line, 3" aperture) and $z=1.0$ (solid dark green line, 1" aperture) as seen in \citealt{carnall2019c} and \citealt{tacchella2021}, with approximately matched quiescent galaxy criteria, to specifically compare the gradient of the age-mass relation seen in simulations with our work. It is interesting to note that the gradient in none of the studies mentioned above (including our work) agree with the $z=1$ snapshot; in fact, our work indicates a steeper gradient in the $z>0.6$ population , and steeper still when we overplot the formation redshifts of the 10 highest redshift cluster galaxies in our sample (black stars, in top right panel). This work agrees with the slope of the $z=0.1$ simulation snapshot, which potentially indicates that for galaxies in our low-z sample ($z<0.95$), either the steep $z=0.1$ relation is already in place, or simulations are not able to reproduce the physical properties of quiescent galaxies at z$\sim$1. Note that the TNG simulations considered here contain both cluster and galaxy-scale halos, but are primarily designed and executed to sample field galaxies (e.g., the small volume of TNG100 does not contain a representative sample of M$_{500c}\sim10^{15}$\Msolar clusters). Dynamically, massive clusters are regions of the Universe with an accelerated clock, and if the star formation in associated halos is similarly accelerated relative to the field, we might expect better agreement in age-mass relations between lower redshift field galaxy simulations and higher redshift cluster observations, as is seen here.

\newcommand{\captiona}{Slopes and Intercepts for Age-Mass relationship}
\newcommand{\arccommentsa}{*Model for the age-mass relation is described by $t_{50,age of Universe}$ (Gyr) = $\alpha$ log$_{10}(M/10^{11}\Msolar) + \beta$ }
\begin{deluxetable*}{l |cc|cc|ccc}
\tablecolumns{7}
\tablewidth{0pt}
\tablecaption{\captiona}
\tablehead{\vspace{-0.2cm} \\ Subsample & \multicolumn{2}{c}{$0.29 < z < 0.45$} & \multicolumn{2}{c}{$0.45 < z < 0.61$} & \multicolumn{2}{c}{$0.61 < z < 0.93$} }
\startdata
& $\alpha$ & $\beta$ & $\alpha$ & $\beta$ & $\alpha$ & $\beta$\\
log(M$_{500c,SZ, z=0}/\Msolar$) $<15$ &  $-1.68\pm0.25$ & $2.66\pm0.09$ & $-1.56\pm0.26$ & $2.69\pm0.09$ & $-1.74\pm0.29$ & $2.09\pm0.10$\\
log(M$_{500c,SZ, z=0}/\Msolar$) $>15$ &  $-1.37\pm0.24$ & $2.62\pm0.08$ & $-1.50\pm0.26$ & $2.54\pm0.08$ & $-2.10\pm0.28$  & $2.37\pm0.09$\\
Early+Mixed Infall           &  $-1.43\pm0.19$ & $2.63\pm0.07$ & $-1.59\pm0.22$ & $2.63\pm0.08$ & $-1.81\pm0.26$ &  $2.14\pm0.09$ \\
Late Infall                  &  $-1.92\pm0.43$ & $2.62\pm0.13$ & $-1.40\pm0.33$ & $2.58\pm0.09$ & $-1.93\pm0.32$ &  $2.39\pm0.10$ \\
\enddata
{\footnotesize \tablecomments{ \arccommentsa}}
\label{table:sed_slopes}
\end{deluxetable*}

\subsubsection{Formation redshifts across M$_{500c,SZ, z=0}$, phase-space location and logM} \label{sec:ks}

In Figure \ref{fig:t50_logm_alldists} (rows 2, 3 and 4), we show distributions of ages of the Universe at galaxy formation redshifts for subpopulations divided by final descendant cluster mass log(M$_{500c,SZ, z=0}/\Msolar$), phase-space location  $p = r_{projected}/r_{500c} \times v_{peculiar}/\sigma_{v}$ (a proxy for infall time), and stellar mass (logM). We conduct this exercise to determine the galaxy property that contributes the most to the difference in formation redshifts observed in our sample: the cluster mass, galaxy stellar mass, or the phase-space location of the galaxy. Vertical lines in each panel correspond to median ages. 

Qualitatively, we find that the galaxy stellar mass subpopulations show the largest difference in formation redshifts, while subpopulations cut by `environmental' factors like cluster subpopulations or phase-space location have similar median formation redshifts. We quantify this observation by using the two-sample Kolmogorov-Smirnov (K-S) test \citep{Hodges1958TheSP}, and obtain the K-S statistic for the following null hypothesis, to check whether the age distributions in each panel in Figure \ref{fig:t50_logm_alldists} are identical or not:\\

\textbf{Hypothesis KS1} : The distributions of formation ages/redshifts — assumed to be \textit{probability distributions} — in each panel are identical (the alternative is that the distributions are not identical).\\

For large samples being considered in a two-sample K-S test, the null hypothesis is rejected at a 5\% significance level if the K-S statistic is $D>D_{threshold}$, where $D_{threshold} = 1.358\times((n+m)/n{\cdot}m)^{1/2}$ \citep{knuth1997}, and $n$ and $m$ are number of elements in the two distributions. For each distribution considered here, $D_{threshold}$ is in the range [0.13,0.19]. 

In Figure \ref{fig:t50_logm_alldists}, we show values of the KS statistic (labeled KS1) in each panel. We see that the null hypothesis is rejected in the higher redshift panels of Row 3 (phase-space) and all panels in Row 4 (stellar mass). We also cannot rule out the hypothesis that the distributions of ages split by log(M$_{500c,SZ, z=0}/\Msolar$) are identical at all redshifts (Row 2).

As expected, we find that there is a significant difference in ages when galaxies are divided by stellar mass logM (Row 4) i.e. high mass galaxies are formed earlier than low mass galaxies. To validate this result, we calculate bootstrapped uncertainties for each age bin (in Row 4), and visually confirm that the distributions of galaxy formation redshifts split by stellar mass are not identical. 

\begin{figure*}[htb!]
\centering
\includegraphics[width=1.02\textwidth]{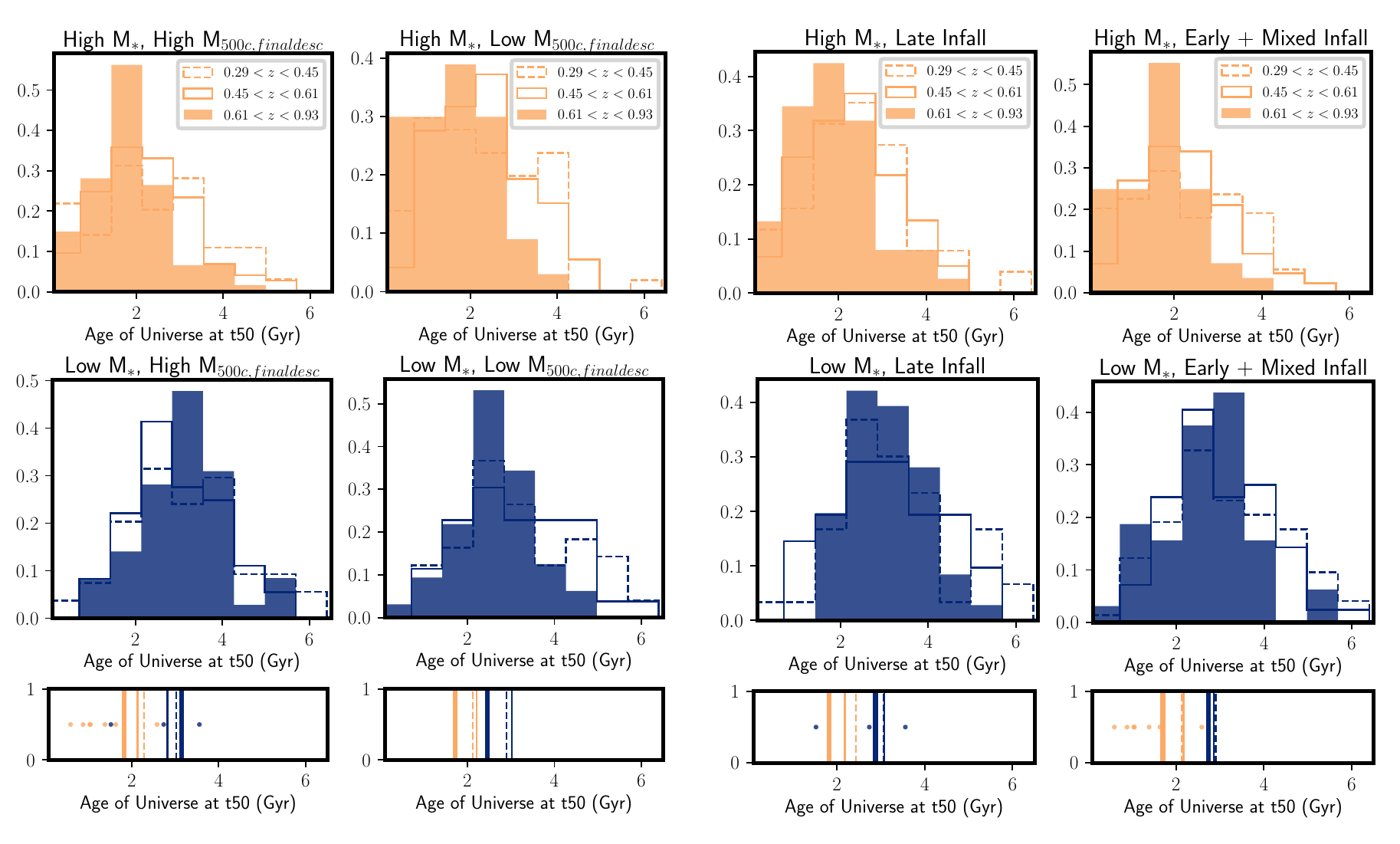}
\caption{Distributions for ages of the Universe at t$_{50}$ (in Gyr) in redshift bins $0.29<z<0.45$ (hollow dashed), $0.45<z<0.61$ (hollow solid), and $0.61<z<0.93$ (filled). Left six panels show distributions for ages where the population is split by log(M$_{500c,SZ, z=0}/\Msolar$), whereas the right six panels show distributions for ages where the population is split by phase-space location, or `infall time'. Stellar mass for each subpopulation is denoted by colors (logM$<10.90$=blue, logM$>10.90$=light orange). The vertical lines in the bottom panels correspond to median ages in each redshift bin (dotted, thin solid and thick solid lines in increasing order of redshift.) Overplotted as dots are the ages of the Universe at t$_{50}$ for the z$>1.2$ galaxies.}
\label{fig:age_t50_hists_1}
\end{figure*}
\GK{The KS statistic value of subpopulation of galaxies in the $0.61<z<0.93$ clusters split by phase-space location implies that a marginal difference in formation redshifts between phase-space subpopulations cannot be ruled out. As discussed in Section \ref{sec:age_lowz} with respect to Figure \ref{fig:age_z_all}, we observe a subpopulation corresponding to a single burst SSP formation redshift of $z>10$ in the early+mixed infall subpopulation. We see this subpopulation (of 27 galaxies) in the low age tails (age of Universe at t$_{50} <$ 2 Gyr) in Row 3 of Figure \ref{fig:t50_logm_alldists} (yellow histogram). When considering the entire distribution of these subpopulations, we find no difference in the median formation age. In cluster environments, we expect galaxies to move along an evolutionary path from late infall to early+mixed infall subpopulations from high to low redshifts. We anticipate this transition to lower the median formation redshifts of the early+mixed infall subpopulation (due to progenitor bias), assuming that the late infall subpopulation has lower median formation redshifts relative to the early+mixed infall subpopulation (as is observed in the highest redshift age distribution). We observe this trend in Figure \ref{fig:t50_logm_alldists} (yellow vertical line, a difference of $\sim$ 1 Gyr across redshift bins). }

\GK{This lack of an accretion history-specific difference in formation age for quiescent galaxies in the two lower redshift bins ($0.29<z<0.61$) is consistent with findings of intermediate redshift cluster studies at $z<0.7$ \citep{sanchezblazquez2009} and SDSS DR7 massive galaxies in \cite{pasquali2019} (where they find that only the lowest mass — logM$<$10 — show sensitivity to cluster environmental effects). If the galaxy cluster is directly affecting the formation of quiescent galaxies (as opposed to being an overall environment in which the `clock' of galaxy evolution runs faster), we would expect to see a signal in the formation redshifts of galaxies tagged as quiescent at that epoch when they are split by accretion history (i.e. phase-space). We only see a hint of this in the highest-redshift ($0.61<z<0.93$) subpopulation. This suggests that even in observably quiescent galaxies, higher redshift clusters are the correct location to observe the echos of cluster-specific transformations that quench star formation \citep{brodwin2013,webb2020}. An in-depth comparison of accretion histories of member galaxies and star formation timescales will be conducted in a future publication. }

As we find stellar mass to be a major contributing attribute in ascribing a formation redshift (or potentially an evolutionary path) to a given galaxy, we inspect each age distribution in Rows 2 and 3 of Figure \ref{fig:t50_logm_alldists} —  divided by log(M$_{500c,SZ, z=0}/\Msolar$) and phase-space location — as a function of stellar mass.

We use the ages plotted in Figure \ref{fig:t50_logm_alldists} to show the age distributions of galaxies in each redshift bin, divided by the two stellar mass bins, to investigate whether the distributions and median ages across redshift for massive quiescent cluster galaxies are drawn from the same parent distribution. See Figure \ref{fig:age_t50_hists_1}, where each panel is a subpopulation split by stellar mass, and environment. This demonstrates that the lower redshift bins have galaxies with extended distributions of formation ages, and the highest redshift bin contains the oldest galaxies. Moreover, the lower mass galaxies in each subpopulation have similar median formation times and distributions. 

In Figure \ref{fig:age_z_all}, we see a median difference in mass-weighted ages between the two stellar mass bins (regardless of redshift bin) of $\sim0.75$ Gyr. In Figure \ref{fig:age_t50_hists_1}, this translates to a median difference of $\sim1$ Gyr (regardless of redshift, final descendant cluster mass, or phase-space location subpopulation). Therefore, this allows us to conclude that the results in this work are consistent with \cite{raichoor2011}, \cite{muzzin2012}, \cite{woodrum2017}, and \cite{webb2020}, that suggest that formation timescales are more varying across stellar mass, and there is only a weak link between formation redshifts for fixed stellar mass across `environment'. 

In Figure \ref{fig:age_t50_hists_1}, we can quantify the lower half (0-50th percentile) of age distributions for a given redshift bin, as the fraction of galaxies formed before the median redshift ($z>3$) in each subpopulation. This metric illustrates that a) \textit{a higher fraction} of more massive galaxy subpopulations forms at $z>3$, compared with low mass galaxies, b) more massive galaxy subpopulations forms on average a Gyr earlier than their corresponding low mass galaxy subpopulation (see bottom panel of Figure \ref{fig:age_t50_hists_1}. We also note a minor difference in formation redshifts between early and late infall galaxies, between 0.2-0.4 Gyr. The highest redshift subpopulations have similar fractions of galaxies with formation redshifts z$>3$, indicating that quiescent galaxies at high-redshifts potentially have similar observational signatures and properties. As galaxies within clusters evolve, and newer systems merge with clusters at lower redshifts, we expect these systems to have ages drawn from wider distributions.

\begin{figure*}[htb!]
\centering
\includegraphics[width=0.75\textwidth]{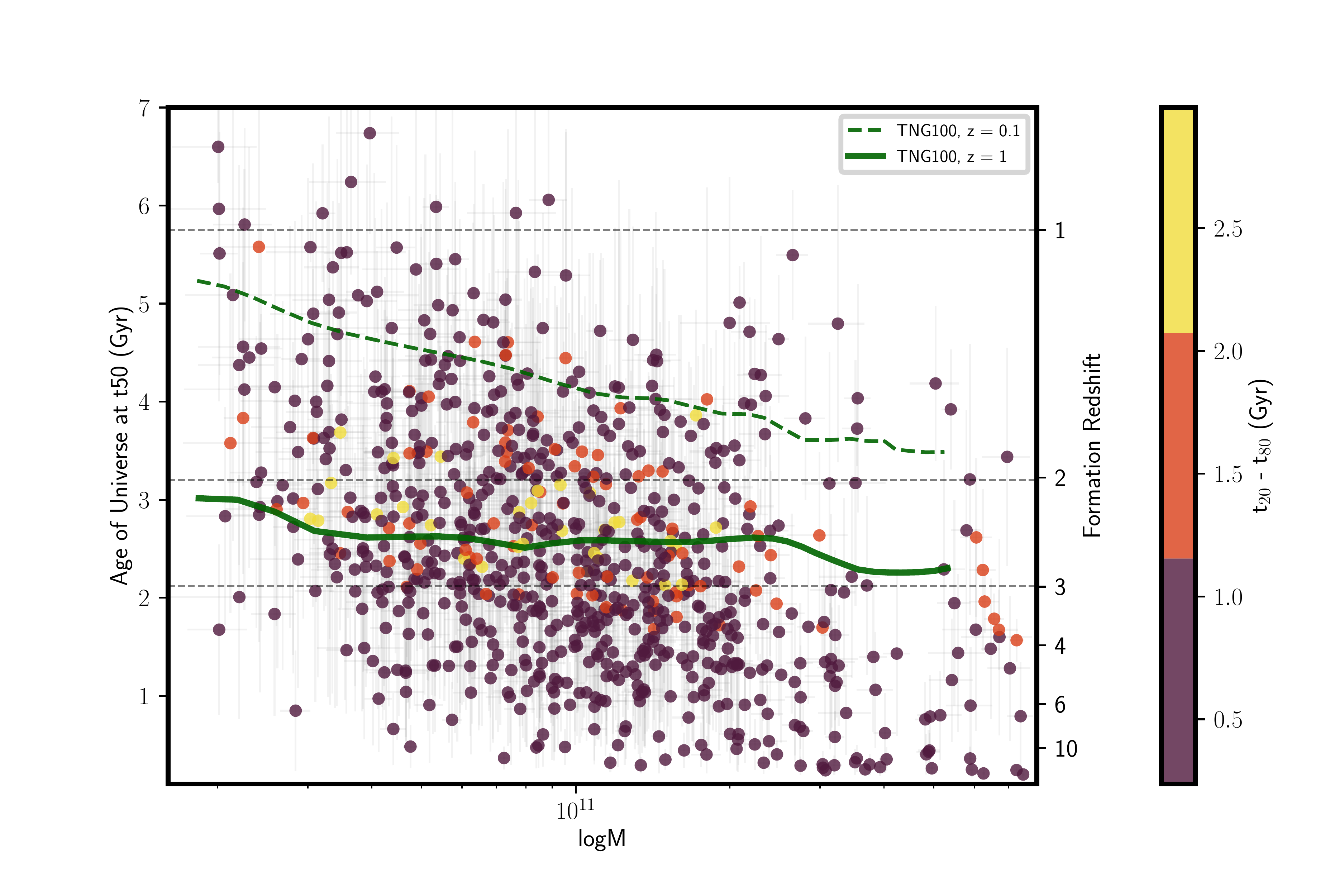}
\caption{Same as Figure \ref{fig:t50_z_mstar}, with points color-coded to represent star formation timescale, as calculated with the parameter t$_{20}$-t$_{80}$ (Gyr). Dark green dotted and solid lines represent ages of Universe at formation times vs log(M) from the TNG100 simulations and redshifts 0.1 and 1 respectively, as seen in Figure 8 of \citealt{carnall2019c}. The most massive galaxies have mostly formed before z$>2$, while the lowest-mass galaxies have formation redshifts of z$<3$. While the most massive galaxies have formed stars in the shortest time-scales, we observe that the most extended star-formation is shown by galaxies at formation redshifts between 1.5-3.5.}
\label{fig:age_z_all_t20_t80}
\end{figure*}

\subsection{Star Formation Timescales and Mass-Dependent Evolution}

Akin to \cite{pacifici2016} and \cite{tacchella2021}, we characterize a notional star formation timescale as the time elapsed between 20\% and 80\% of stellar mass formed for a given galaxy, t$_{20}$-t$_{80}$ (in units of Gyr). In Figure \ref{fig:age_z_all_t20_t80}, we plot t$_{50}$ vs logM, with colors indicating the star formation timescale. \GK{We find that the most massive galaxies on average only exhibit shorter star formation timescales (i.e. for logM $>$ 11.5, the median star formation timescale is 0.45$^{+0.08}_{-0.07}$ Gyr, whereas median timescales for the logM $<$ 11.5 is 0.64$^{+0.04}_{-0.02}$ Gyr)}; this is consistent with other studies (see references below). This analysis has its shortcomings; see Appendix D in \cite{tacchella2021} for comparisons between parametric and non-parametric SFH vis-a-vis prior imprints on calculations of timescales (such as quenching and star-formation timescales). They find that galaxies with formation redshifts $z<3$ have longer star formation timescales, but our analysis interestingly only reproduces that trend for a subset of galaxies with z$_{formation}$ between $1.5 < z < 3.5$ for timescales $> 1$ Gyr.

The majority of star formation in massive quiescent galaxies (most of which morphologically look like early-type galaxies) occurs at high redshifts, with passive evolution thereafter (see Section 1, and \citealt{vandokkum1998,jorgensen2006,saracco2020,tacchella2021}). \cite{sanchezblazquez2009} study stellar populations in red-sequence galaxies in clusters and groups at $0.4<z<0.8$ and measure formation redshifts of $z>2$, and find that those massive galaxies are compatible with passive evolution since. \cite{sanchezblazquez2009}, \cite{gallazzi2014}, and \cite{webb2020} also find that the most massive galaxies in their datasets form stellar mass earlier and quicker, relative to lower mass systems; higher-redshift studies also point towards this trend (see Section 5.1 on mass-dependent evolution in \cite{webb2020} and Section 6.2 in \cite{diazgarcia2019} for more information, and references therein). 

Our objective is to quantify star-formation timescales in quiescent galaxies. We observe the above mass-dependent evolution in our studies, where formation redshifts between high and low-mass systems differs substantially. Studies have attempted to explain this mass-dependent evolution either by accounting for the different methods to calculate metallicity, or due to the parameterization (or lack thereof) of SFHs, that can potentially bias formation redshifts (see Section \ref{sec:results}). t$_{20}$-t$_{80}$ is a simple parameterization to achieve this goal; a given value of t$_{20}$-t$_{80}$ does not correspond to a unique shape of the SFH. Non-parametric SFHs could be key here. We will conduct an analysis of this dataset with non-parametric SFH models to measure quenching timescales in a follow-up work (Khullar et al., in prep).

\begin{figure}[htb!]
\centering
\hspace{-0.6cm}%
\includegraphics[width=0.5\textwidth]{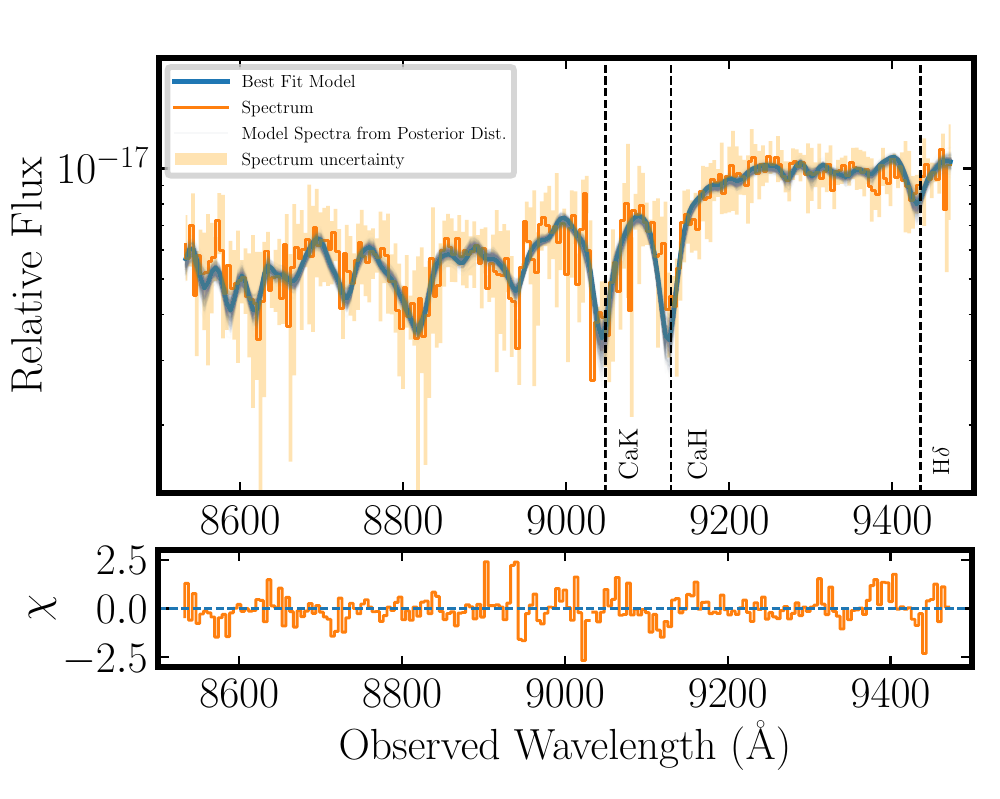}
\caption{(Top) Stacked spectrum (orange) and best-fit SED models (blue) for 10 galaxies in the  z$>1.2$, log(M$_{500c,SZ, z=0}/\Msolar$)$>15$ bin.}
\label{fig:stack_hiz}
\end{figure}

\subsection{Formation of galaxies observed at $z>1.2$}\label{sec:dedicated_hiz}

In the bottom panels of Figure \ref{fig:age_t50_hists_1}, we overplot as dots the ages of the Universe at t$_{50}$ for 10 massive cluster quiescent galaxies at $z>1.2$ in our sample (z$_{median}$=1.3). All 10 galaxies belong to the log(M$_{500c,SZ, z=0}/\Msolar$)$>15$ subpopulation. The median age of the Universe at t$_{50}$ calculated via a stacked spectrum (see Figure \ref{fig:stack_hiz}) SED fit for this subpopulation is $1.42\pm0.74$ Gyr, corresponding to a formation redshift of $4.33_{-1.31}^{+3.38}$ (a 1$\sigma$ range of $z=3-7$).

\GK{Conducting a direct comparison with recent field galaxy studies}, we find that the formation redshifts of galaxies in our $z>1.2$ sample are either similar or marginally higher. By determining SFHs for 75 massive quiescent field galaxies at $1<z<1.3$ with a median stellar mass of logM$\sim$11, \cite{carnall2019} find a mean formation redshift of 2.6, with a range of formation redshifts between 1.5-6. \cite{tacchella2021} measure quenching timescales and SFHs for 161 galaxies, with $\sim$20 galaxies at $z>1$ and an aggregate formation redshift of $~4$, consistent with this work. \cite{saracco2009} find that their study of 32 quiescent early-type galaxies (ETGs) at z$\sim$1.5 divides them into young and old systems, with the older population forming the bulk of their stars between redshifts z$\sim$5-6, while the younger galaxies form at z$\sim$2-3. While our sample has a small number of $z>1.2$ quiescent galaxies, Figure \ref{fig:t50_logm_alldists} indicates that our highest-redshift high mass galaxies (logM>11) form at $4<z<10$.

When considering cluster galaxies, \cite{raichoor2011} measure ages and stellar masses for 79 cluster ETGs at z$\sim$1.3 (albeit with multi-band photometry and \cite{bruzual2003} SED models); for galaxies with logM$>$10.5, they find formation redshifts of $2<z<10$, marginally wider than the distribution in this study. For 331 quiescent galaxies in galaxy clusters at $1<z<1.4$, \cite{webb2020} sample a similar range of masses as our study, and find that the majority of the highest stellar mass galaxies have an aggregate formation redshift of z$\sim5.4$ (logM$>11.3$, with a range of $z=3-10$), while lower mass galaxies have a formation redshift of z$\sim3.3$ (logM$<10.5$, with a range of $z=2-8$); our results (both the median and range) for the $z>1.2$ sample are in agreement.

\section{Challenges and Future Work} \label{sec:challenges}

Comparing ages of stellar populations in massive and quiescent galaxies across cosmic time in various studies is a non-trivial task, especially due to the fact that ages across studies are calculated via different methods and modeling techniques. Moreover, the impact of metallicity is crucial, and the extent to which age-metallicity degeneracy is broken in this work needs to be investigated further, by measuring other age and metallicity indicators, especially via direct absorption line measurements (e.g., \citealt{choi2014}, and see Appendix in \citealt{webb2020}). Finally, it should be noted that most studies of massive galaxies use UVJ color-based selection to select quiescent galaxies, which is an approach we did not utilise. 

Further photometry and spectroscopy in the infrared would allow us to characterize properties of dust-unobscured stellar populations in these systems. Datasets like the just-completed SPT-HST SNAP cluster imaging of $137$ SPT clusters at $0.3<z<1.5$ with F110W and F200LP photometry (Remolina-Gonzalez et al. in prep) will allow us to morphologically characterize the brightest galaxies in these systems as well (see examples of such analyses in \citealt{belli2015,estradacarpenter2020,Akhshik2020,matharu2020}).  We also note that a study of cluster mass accretion histories in simulations could highlight the (possibly non-negligible) population of `pre-processed' quiescent galaxies in our sample i.e. galaxy group environments that could cause infall-based quenching of galaxies before they enter accrete to their final cluster halo \cite{1998ApJ...496...39Z,2019MNRAS.488..847P}. 

The ability of delayed-tau SFH models to constrain quenching timescales has been called into question \citep{carnall2019,leja2019}. A modification of the current methodology that will be explored in future work (Khullar et al., in prep) is the usage of non-parametric SFHs, and by using frameworks that constrain star formation episodes in SFHs via the dense basis method \citep{iyer2019}. We will also explore mass-weighted ages with calculations of mass accretion histories of cluster haloes studied in simulations (e.g., IllustrisTNG, \citealt{pillepich2018}).  

\section{Summary}
\label{sec:conc}

In this work, we characterize stellar populations in massive cluster quiescent galaxies from the SPT-GMOS survey \citep{bayliss2016} and the SPT Hi-z survey \citep{khullar2019}, to constrain stellar masses, ages and SFHs in 837 galaxies at $0.3<z<1.5$. We constrain these properties via SED analysis of individual systems' photometry and optical spectroscopy, with the Bayesian fitting framework \code{Prospector} and primarily a delayed-tau SFH model. We calculate mass-weighted ages and formation redshifts for galaxies as a function of stellar mass to quantify mass evolution with time. We measure formation redshifts in different environments; `environment' in this work is characterized by placing galaxies in subpopulations divided by final descendant galaxy cluster mass M$_{final,desc}$, and phase space location — a proxy for infall time — r$_{projected}/r_{500c}$ x $v_{peculiar}/\sigma_{v}$). We also employ stacked spectra to robustly characterize aggregate properties of the the highest redshift galaxies with low SNR spectra and boost wavelength coverage, as well as to cross-check our analyses of median properties. We find that:

\begin{itemize}
    \item Quiescent galaxies in our dataset sample a diverse set of SFHs, exhibiting a range of mass-weighted ages as a function of redshift, and environment — with 6.23$_{-1.38}^{+1.41}$ Gyr being the 16th, 50th and 84th percentile age distribution (median uncertainty of 1.22 Gyr). 
    \item The median formation redshift in our sample is {$2.8\pm0.5$}, with a range of $z=1-6$, and is on aggregate similar or marginally older than massive quiscent field galaxy studies, and similar to cluster studies at $z>1$. On average, we find that more massive galaxies form $\sim$0.75 Gyr earlier than lower mass galaxies.
    \item The highest redshift galaxies in our sample ($z>0.6$) show a marginally steeper age-mass relation relative to lower redshift subpopulations, indicating that the age-mass relation does not change (within uncertainties) at ($z<0.6$) in our cluster quiescent galaxies sample.
    \item The median age-mass relation (slopes and intercepts) of the full sample is t$_{50,age of Universe}$ = $2.52(\pm0.04) - 1.66(\pm0.12)$ log$_{10}(M/10^{11}\Msolar)$, similar to other massive field quiescent galaxy studies seen in the literature.
    \item Lower mass quiescent galaxy subpopulations across M$_{final,desc}$ and phase-space location form approximately at the same formation redshifts ($z\sim$2), regardless of the observed redshift bin.
    \item Subpopulations that have interacted the most with their respective galaxy cluster's gravitational potential i.e. log(M$_{500c,SZ, z=0}/\Msolar$)$>15$ and (r$_{projected}/r_{500c}\times v_{peculiar}/\sigma_{v}) <$ 0.4 (early infall time) have steeper age-mass relations relative to other subpopulations, indicating marginal influence of environmental quenching.
    
    \end{itemize}
    
This is the first publication in a series which will enable studies of stellar mass assembly in clusters across a wide range of redshifts. With upcoming spectroscopic datasets of clusters at $z>1$, we will comprehensively determine star formation and quenching timescales in quiescent galaxies, and connect galaxies at high redshifts to lower-redshift objects in an antecedent-descendent manner. 

\section{Acknowledgments}

GK thanks Tom Crawford, Ben Johnson, Joel Leja, Sandro Tacchella, Kate Whitaker, Nora Shipp and Lucas Secco for their feedback on the methods and analysis shown in this paper. The authors would like to express gratitude towards the staff and workers at the 6.5m Magellan Telescopes at the Las Campanas Observatory, and the Gemini-South telescope in Chile, for their valuable labor. 

The South Pole Telescope program is supported by the National Science Foundation (NSF) through grants PLR-1248097 and OPP-1852617. Partial support is also provided by the NSF Physics Frontier Center grant PHY- 1125897 to the Kavli Institute of Cosmological Physics at the University of Chicago, the Kavli Foun- dation, and the Gordon and Betty Moore Foundation through grant GBMF\#947 to the University of Chicago. Argonne National Laboratory’s work was supported by the U.S. Department of Energy, Office of Science, Of- fice of High Energy Physics, under contract DE-AC02- 06CH11357. The University of Melbourne authors acknowledge support from the Australian Research Council’s Discovery Projects scheme. GM received funding from the European Union’s Horizon 2020 research and innovation programme under the Marie Skłodowska-Curie grant agreement No MARACAS - DLV-896778. AS is supported by the ERC-StG ‘ClustersXCosmo’ grant agreement 716762, by the FARE-MIUR grant 'ClustersXEuclid' R165SBKTMA, and by INFN InDark Grant.

\facilities{Magellan Telescopes 6.5m (Clay/LDSS3C, Clay/PISCO, Baade/FOURSTAR), Gemini-North Telescope (GMOS)}

\software{\code{Python 3.6 — Prospector, python-FSPS, SEDpy, pygtc, Matplotlib, Numpy, Scipy, Astropy, Jupyter, IPython} Notebooks, GALFIT, SAO Image DS9, IRAF, IDL}

\bibliographystyle{aasjournal}
\bibliography{main}

\appendix

\section{Metallicity: log($Z/Z_{\odot}$)}

\begin{figure*}[htb!]
\centering
\includegraphics[width=0.7\textwidth]{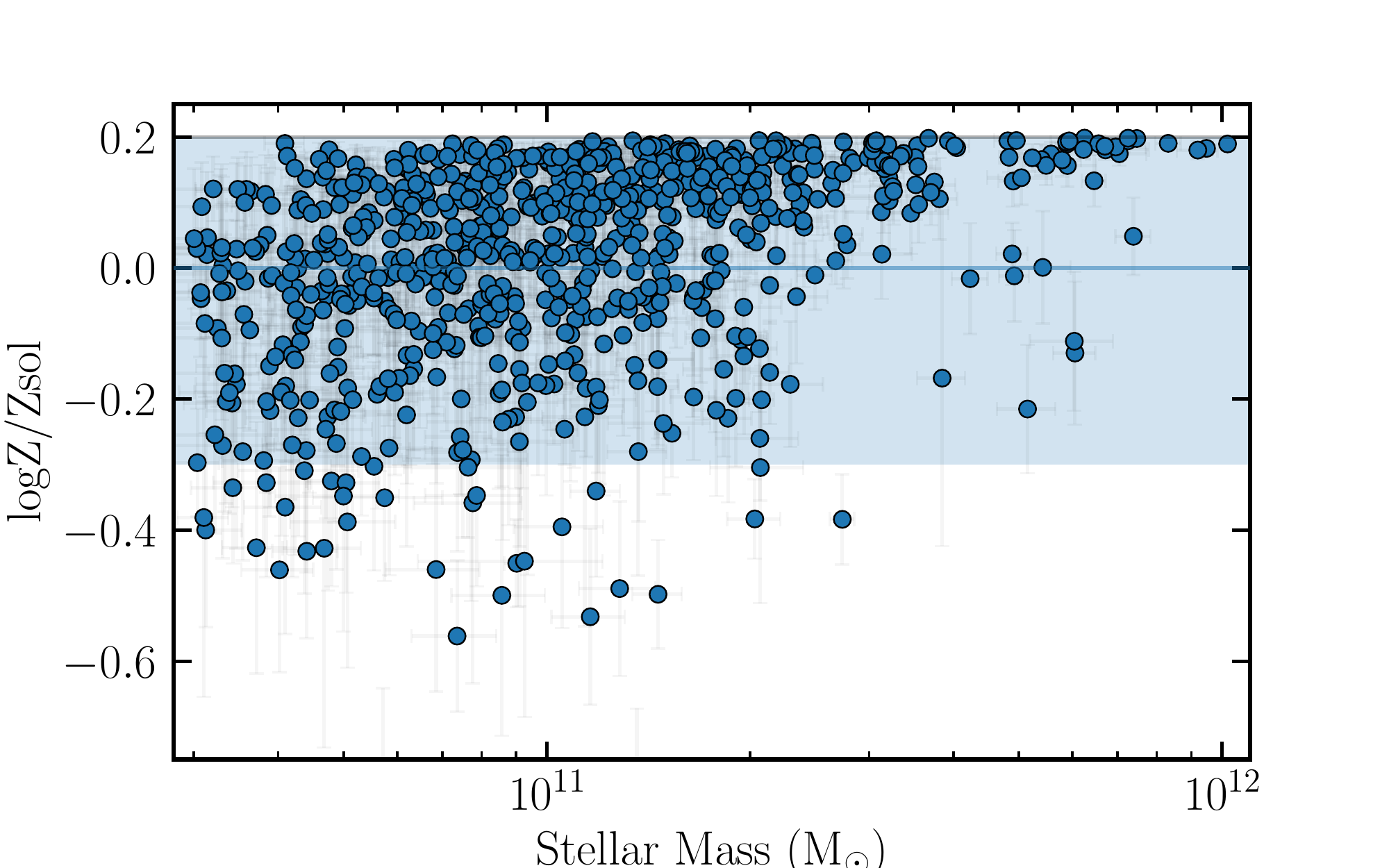}
\caption{Metallicity distribution for galaxies considered in this work, as a function of stellar mass. The blue horizontal line corresponds to the mean metallicity assigned to a clipped normal prior for each SED model fit, with the shaded region corresponding to the 1$\sigma$ prior range bound on the upper end at log(Z/Z$_{\odot}$)=0.}
\label{fig:met}
\end{figure*}

Metallicity for individual galaxies in our SED analysis is fit as a free parameter within the \code{Prospector} framework. In Model A, we fit for metallicity by using the mass-metallicity relation (MZR) from \cite{Gallazzi2005} as a starting point, and incorporate studies about the evolution of stellar mass-metallicity relation in clusters \citep{ellison_massmetallicity_2009,Leethochawalit2018} such that for each individual galaxy fit, we use a clipped-normal prior centered at log(Z/Z$_{\odot}$)=0.0, with a dispersion of 0.3, clipped at [-2.0,0.2]; the bounds are defined by MIST and MILES libraries used in \code{Prospector}. With optical spectroscopy, we rely on spectral signatures in the rest-frame 3710-4120\AA\ range to break the age-metallicity degeneracy.

Figure \ref{fig:met} shows the median metallicities log(Z/Z$_{\odot}$) as a function of stellar mass (simultaneously fit with metallicity) from Model A fits.  We find that the highest mass galaxies (logM$>11$) have median metallicities in a narrow range, while low-mass galaxies have a diverse set of median metallicities. This result has an impact on the creation of stacked spectra, as care is needed to assign and fit metallicity in a stacked spectrum, especially when the constituent individual spectra span a wide range of metallicities (variation in log(Z/Z$_{\odot}$) does not scale linearly with flux).

\section{Age biases in median stacking of spectra}

\begin{figure*}[htb!]
\centering
\includegraphics[width=0.85\textwidth]{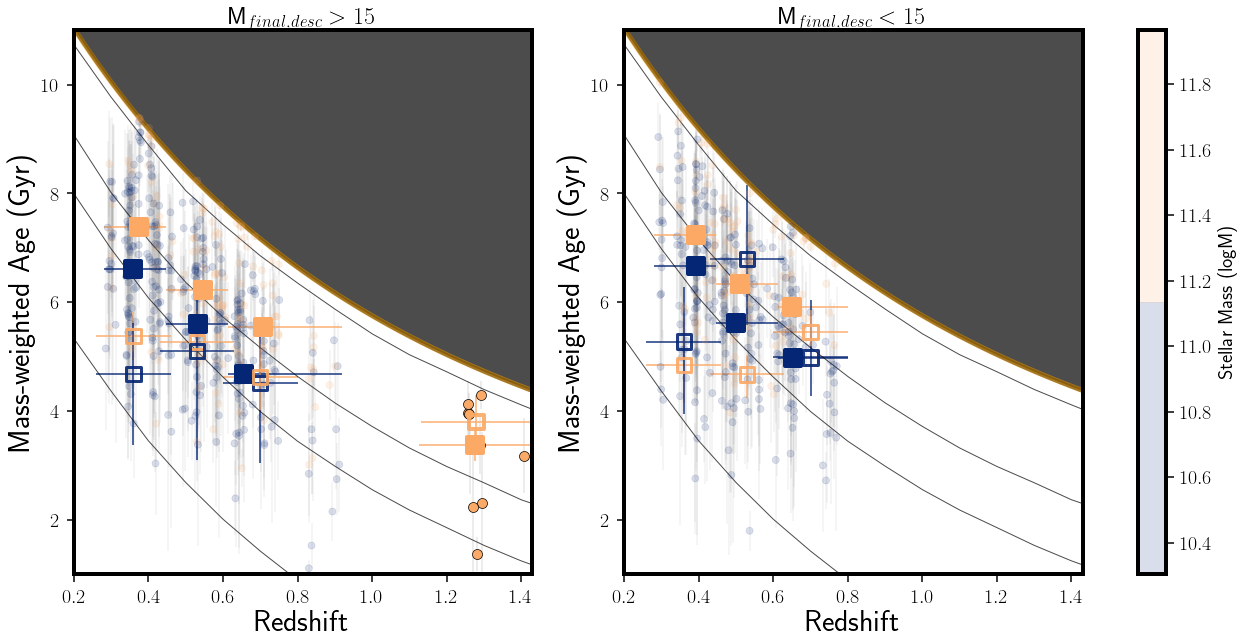}
\caption{Same as top panels of Figure \ref{fig:analysis_and_stack}, but hollow squares signifying stacked spectra mass-weighted ages where stacking is performed using the alternate method described here.}
\label{fig:stack_bias}
\end{figure*}

As mentioned in Section 4.2, we explore an alternate method of stacking, where uncertainty per flux element is characterized by calculating the uncertainty on the median flux (from median fluxes per wavelength element in a given stack bin); this is the usual approach to stacking seen in SED studies, to visually qualify and quantify spectral features and galaxy properties. We find that this method severely underestimates uncertainty, and generates median mass-weighted ages that are biased by $\sim 1.5$ Gyr ($\sim 0.8$ Gyr) in higher (lower) stellar mass stacks in the lowest redshift bins. See Figure \ref{fig:stack_bias}, which plots stack ages as a function of redshift for galaxies divided by log(M$_{500c,SZ, z=0}/\Msolar$) subpopulations, and compares them with median ages in a given bin.

\begin{figure*}[htb!]
\centering
\includegraphics[width=1.0\textwidth]{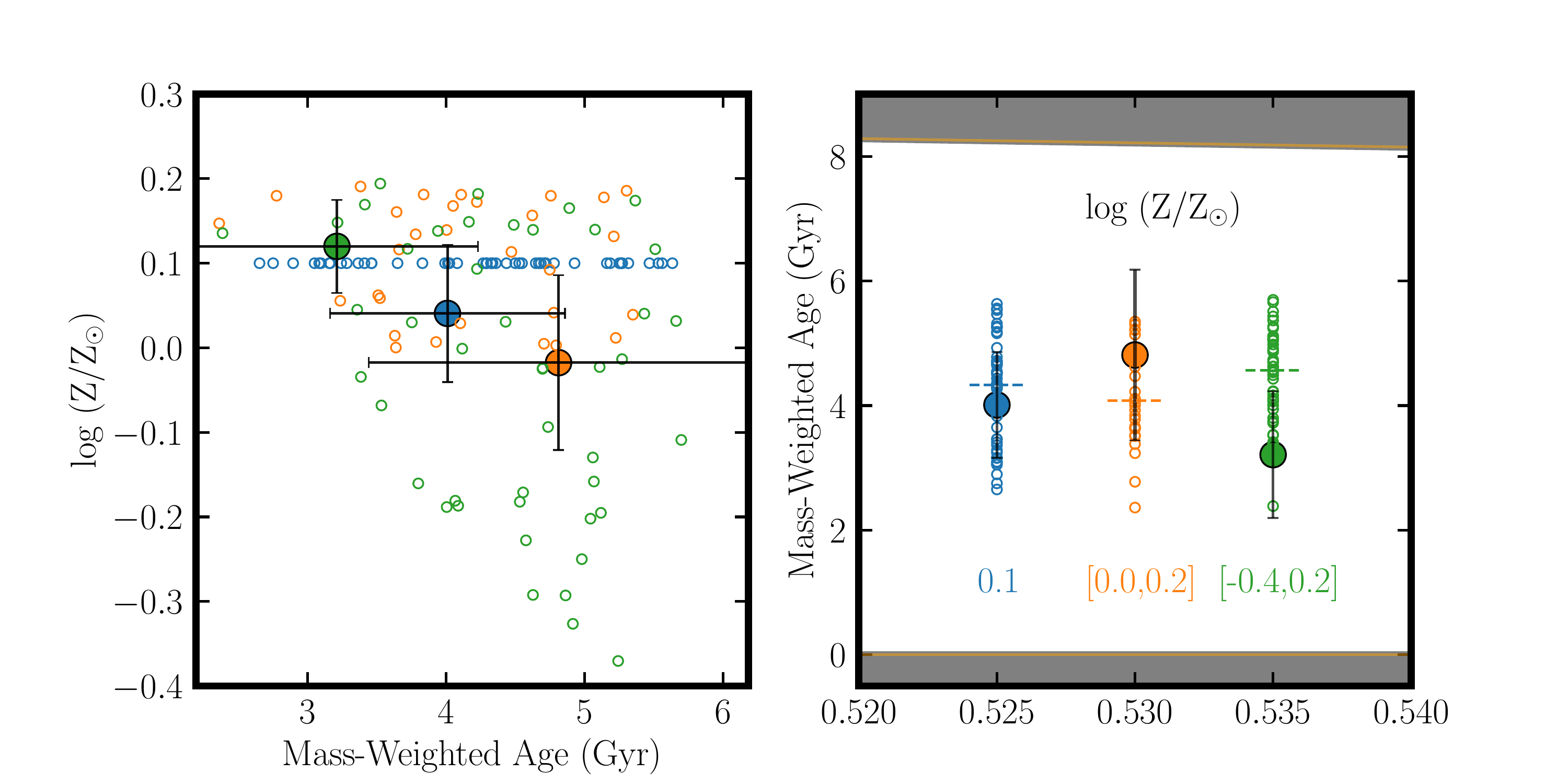}
\caption{Physical properties of three mock samples of quiescent galaxies: Sample 1 (fixed metallicity log(Z/Z$_{\odot}$) = 0.1, in blue), Sample 2 (Metallicity in a restricted range log(Z/Z$_{\odot}$) = [0.0,0.2], in orange) and Sample 3 (Metallicity in a restricted range log(Z/Z$_{\odot}$) = [-0.4,0.2], in green). (Left) Metallicity vs mass-weighted age for stacked spectra from three samples with varying metallicities (filled circles) and individual galaxy per sample (empty circles). (Right) Mass-weighted ages for the three samples, with median ages for each sample annotated with horizontal dotted lines.}
\label{fig:stack_bias2}
\end{figure*}
\begin{figure*}[htb!]
\centering
\includegraphics[width=0.85\textwidth]{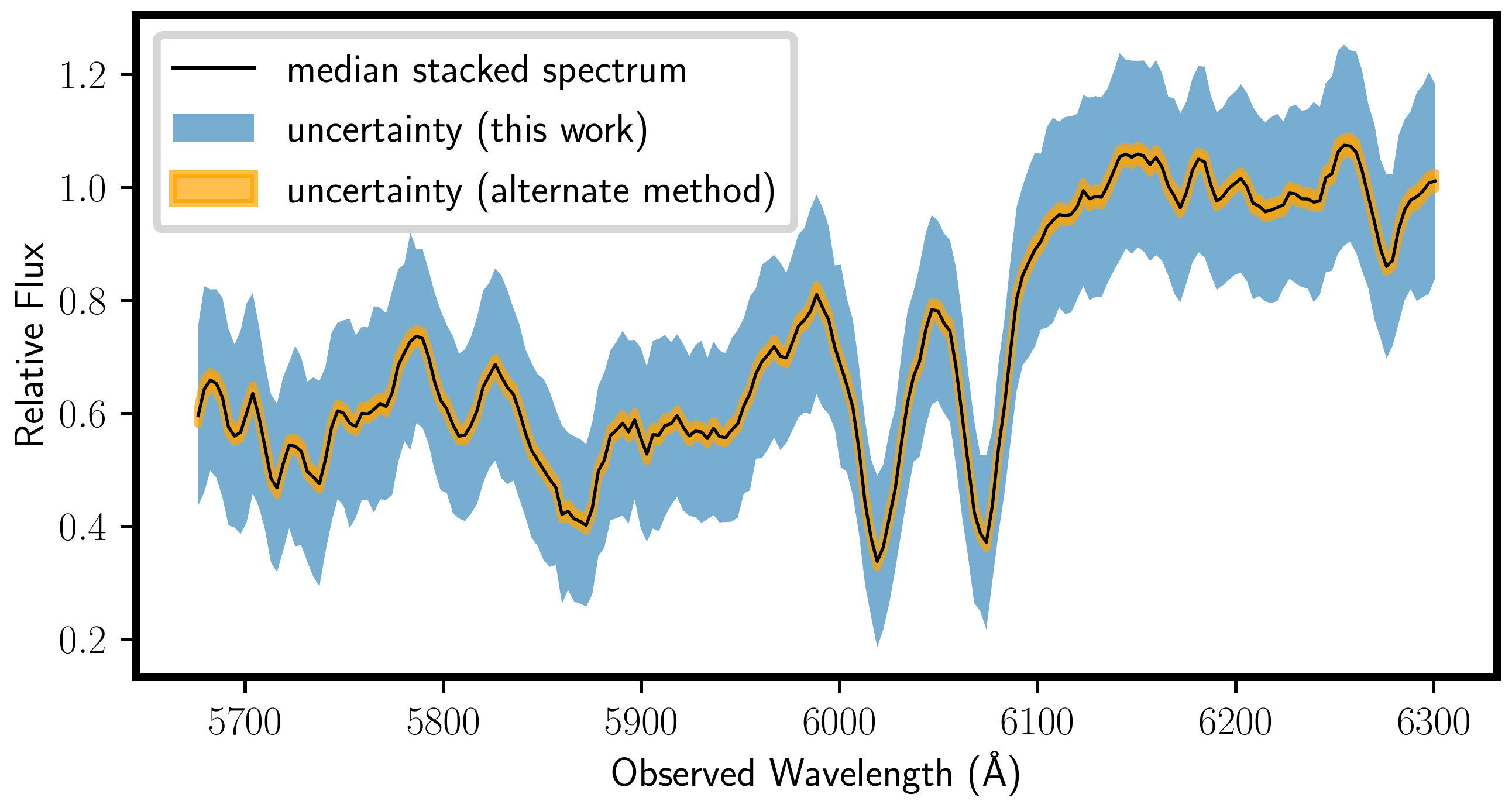}
\caption{Stacked spectrum generated for galaxies in the log(M$_{500c,SZ, z=0}/\Msolar$)$>15$, logM$>10.90$ and $z=0.53$ bin in our sample. The median stacked spectrum is plotted in black, while the stacked spectrum uncertainty considered in this work is plotted in blue. Orange denotes the uncertainty derived from the alternate stacking method, which — as we argue in this section — biases age calculations and may underestimate uncertainty.}
\label{fig:stack_mock_2unc}
\end{figure*}

To quantify this bias, we generate stacks via this alternate method for a given redshift and stellar mass bin, for mock galaxies. We do this by sampling galaxy SEDs via \code{Prospector} from the allowed parameter space for the log(M$_{500c,SZ, z=0}/\Msolar$)$>15$, logM$>10.90$ and $z=0.53$ bin galaxies, using Model A (see Section 4.2). This corresponds to an age range of [0,8] Gyr, and a stellar mass range of logM=[10.90,12.0].

We make three samples, with each galaxy samplying varying range of metallicities:\\
1. Sample 1: fixed metallicity log(Z/Z$_{\odot}$) = 0.1.\\
2. Sample 2: Metallicity in a restricted range log(Z/Z$_{\odot}$) = [0.0,0.2] (range observed in our highest mass galaxies)\\
3. Sample 3: Metallicity in a restricted range log(Z/Z$_{\odot}$) = [-0.4,0.2] (range observed in our lowest mass galaxies). \\

We make a D4000 $>1.45$ cut on the sampled SEDs, with an average of 40 galaxies in each sample. We use similar priors on all parameters as Model A, while the metallicity prior is approximately 2$\sigma$ times the priors from Model A. We pass these stacked spectra through a similar analysis as is conducted in this work. 

Figure \ref{fig:stack_bias2} shows the distribution of stacked metallicities and ages for each sample, with filled points corresponding to stack values, and hollow points corresponding to parameter values for the individual mock galaxies in each sample. Horizontal dotted lines correspond to median mass-weighted ages per sample. 

We find that the stack metallicity and age is the most biased for Sample 3, with the highest range in metallicity, while Sample 1 is the least biased i.e. for fixed metallicity, we find that the stacked spectra retrieves ages matching median age of the sample of constituent galaxies. In Sample 3, we see a bias as wide as $1.5\sigma$ (in this specific case, an age that is younger than the median age by $\sim1.5$ Gyr). Hence, we attribute that the dominant source of the bias in ages from this stacking method, is the range of metallicities in the constituent galaxies per stacking bin. This is a bigger contributing factor in stacks from the lower mass galaxies, since these subpopulations are where we see the largest range in metallicities. Hence, we do not employ this stacking method in this work. 

\section{Single Burst SFHs}

\begin{figure*}[htb!]
\centering
\includegraphics[width=1.0\textwidth]{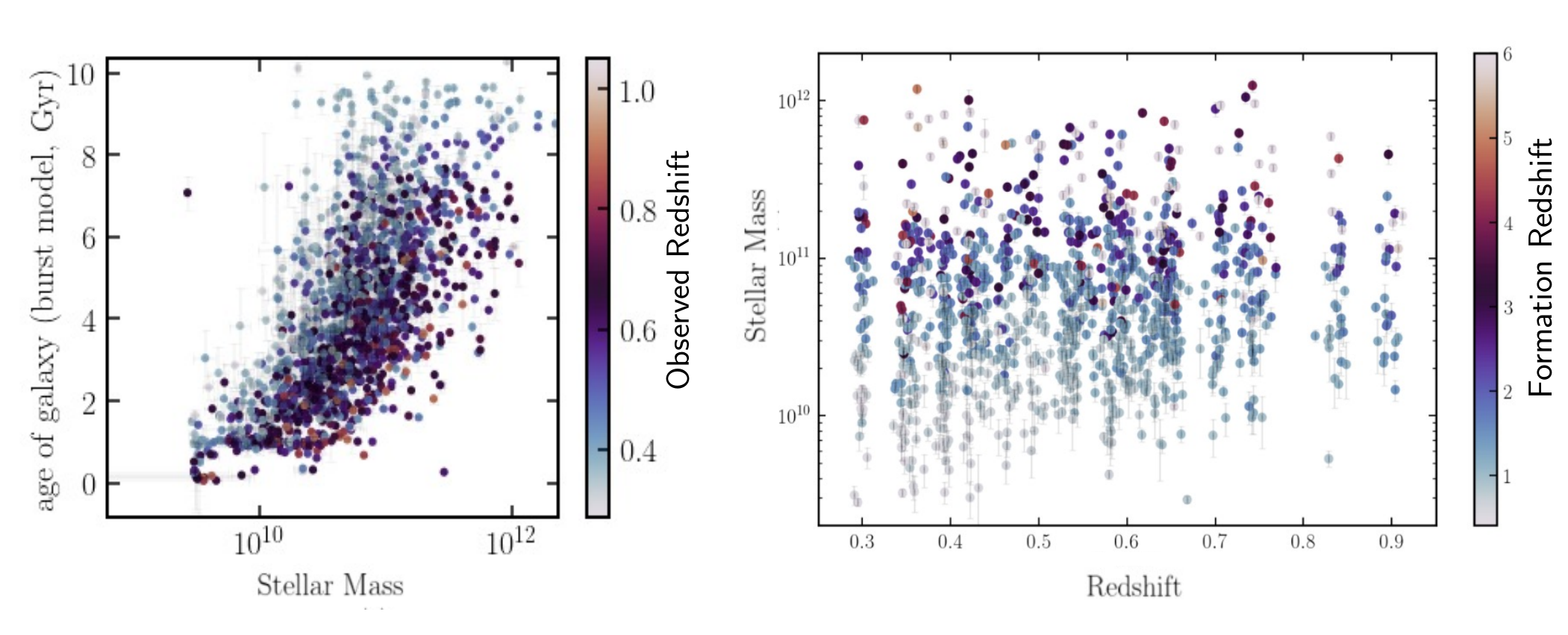}
\caption{(Left) Ages (in Gyr) as a function of stellar mass (in \Msolar) for a single burst fixed metallicity model-based (Model B), with color indicating observed redshifts. (Right) Stellar Mass vs redshift for the low-z member galaxies in our sample, corresponding to Model B. Colors correspond to formation redshifts, where more massive galaxies (logM $> 11$) were formed at z$>3$. }
\label{fig:burst_age}
\end{figure*}

Beyond the delayed-tau SFHs, we also calculate the age/epoch of star formation (in the form of a single burst-like star formation age) for the single burst model (Model B). This is a more limiting model for galaxies with more than one episode of star formation (which would better be approximated by the delayed-tau SFH), but is an exercise to test the robustness of approximating quiescent galaxies as simple stellar populations, a model employed by many studies in the past (e.g., \cite{fumagalli2016,jorgensen2017}. See Figure \ref{fig:burst_age} for stellar age of quiescent galaxies as a function of stellar mass (M) for galaxies in our sample at $0.3<z<0.9$. Color of points in the figure indicates observed redshift of the member galaxy. As is expected, the most massive galaxies are formed earliest, with a median-age of $\sim$ 5 Gyr for a massive quiescent cluster member galaxy in our sample.

As expected, objects observed at lower redshifts have older ages i.e. for a given stellar mass, low redshift galaxies sit on the top end of the plot. To physically motivate this, and compare this distribution of ages to the evolution of galaxies in the Universe, we map these ages and observed redshifts to the formation redshift of each galaxy for a distribution of stellar mass M as a function of observed redshift, where color indicates the formation redshift of a galaxy z$_{form}$). The median mass galaxy (M) has formed at z$_{form}$ = 2-3, given the assumptions of a single burst model; this is significantly younger than results seen from Model B, as is expected in an SSP-model based age characterization.

\section{SNR of quiescent galaxy spectra}

\begin{figure*}[htb!]
\centering
\includegraphics[width=0.9\textwidth]{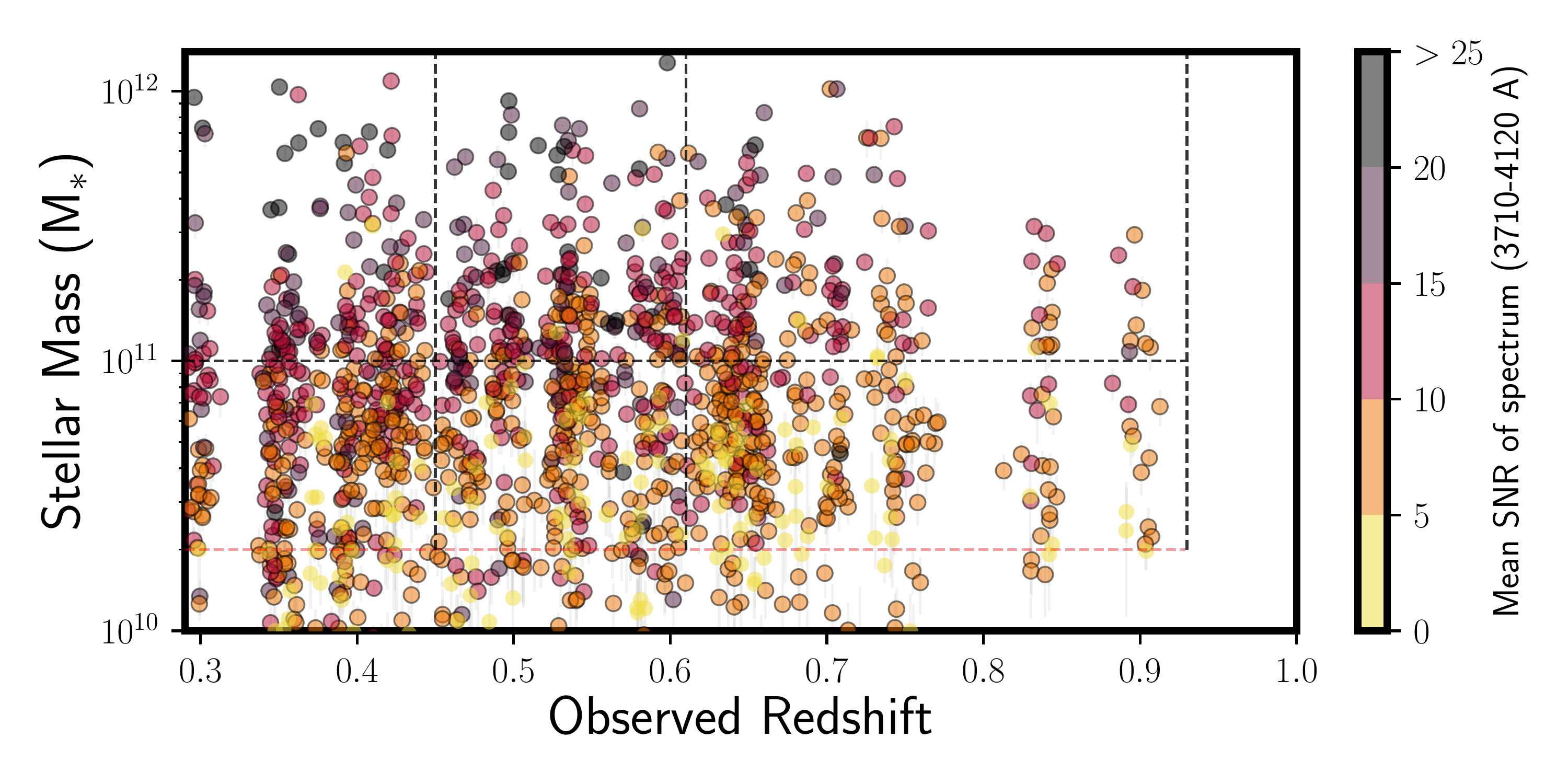}
\caption{Same as Figure \ref{fig:mstar_dtau_met}, with average signal-to-noise ratio of the observed spectrum coded with color for each galaxy spectrum in the low-z cluster sample ($0.3<z<0.9$), where we employ an SNR cut. Points with black borders are the 827 galaxies considered in this study from the low-z cluster sample. The highest SNR spectra were observed from the higher mass galaxies in the sample.}
\label{fig:snr}
\end{figure*}

We apply a mean spectrum SNR cut to our quiescent galaxy spectra in the range $0.3<z<0.9$. Figure \ref{fig:snr} shows the distribution of stellar mass as a function of observed redshift, with color indicating mean SNR per galaxy spectrum. 

We find that the highest SNR spectra are observed in the highest mass galaxies, which is expected, without a strong redshift dependence, as is expected from observational design of the program in B16. Intermediate mass galaxies are seen to have been derived from a flat distribution of intermediate SNR spectra, mostly independent of redshift in the low-z sample. The SNR distribution also indicates that the lowest SNR ($<5$) galaxies are cut from the sample by applying the mass cut (logM $>10.3$, dotted red lines in Figure \ref{fig:snr}).

\end{document}